\documentclass{aa}  

\usepackage{graphicx}
\usepackage{txfonts}

\usepackage{xcolor}

\usepackage{amstext}

\begin{document}

   \title{Multiwavelength observations of Swift J0243.6+6124 \\from 2017 to 2022}

   \subtitle{}

   \author{Wei Liu\inst{1,2}
          \and
          Jingzhi Yan\inst{1} 
          \and
          Pablo Reig\inst{3,4}
          \and
          Xiaofeng Wang\inst{5,6}
          \and
          Guangcheng Xiao\inst{1}
          \and
          Han Lin\inst{5}
          \and
          Xinhan Zhang\inst{5}
          \and
          Hanna Sai\inst{5}
          \and
          Zhihao Chen\inst{5}
          \and
          Shengyu Yan\inst{5}
          \and
          Qingzhong Liu\inst{1}
          }

   \institute{Key Laboratory of Dark Matter and Space Astronomy, Purple Mountain Observatory, Chinese Academy of Sciences, \\Nanjing, 210023, China\\
              \email{weiliu@pmo.ac.cn}, {jzyan@pmo.ac.cn}
         \and
             School of Astronomy and Space Science, University of Science and Technology of China, Hefei, 230026, China
        \and
            Institute of Astrophysics, Foundation for Research and Technology-Hellas, 71110 Heraklion, Greece
        \and
            Physics Department, University of Crete, 71003 Heraklion, Greece
        \and
            Physics Department/Tsinghua Center for Astrophysics, Tsinghua University, Beijing, 100084, China
        \and
            Beijing Planetarium, Beijing Academy of Sciences and Technology, Beijing, 100044, China
             }

\date{Received 27 May 2022 /
       Accepted 12 July 2022}
       
 
  \abstract
    {Swift J0243.6+6124 is a high-mass X-ray binary that went into a giant X-ray outburst in 2017. During this event, the X-ray luminosity reached the highest value ever measured in a galactic Be/X-ray binary.}
    {Our aim is to study the long-term variability of Swift J0243.6+6124 after the 2017 major X-ray outburst.}
    {We have obtained optical spectroscopy and photometry data during four years after the event. The long-term photometric light curve and the equivalent widths of the H$\alpha$ and He I $\lambda$6678 lines were used to monitor the state of the Be star's circumstellar disk. The H$\alpha$ line profiles show evidence for V\,/\,R variability that was accounted for by fitting the H$\alpha$ spectral line profile with two Gaussian functions.
    We divided our data into three phases according to the intensity of the X-ray, optical, and infrared emission.}
    {Phase I covers the rise and decay of the giant X-ray outburst that took place in October--November 2017. We interpret phase II as the dissipation of the Be star's equatorial disk and phase III as its recovery. The timescale of a complete formation and dissipation process is about 1250 days. 
    The epoch when the dissipation process stopped and the reformation period began is estimated to be around MJD 58530. We find a delay of $\sim$\,100\,--\,200 days between the minimum of the optical or infrared intensity and the strength of the H$\alpha$ line after the X-ray outburst, which may indicate that the dissipation of the disk begins from the inner parts.
    The motion of the density perturbation inside the disk is prograde, with a V\,/\,R quasi-period of $\text{about four}$\, years.
    The source shows a positive correlation in the $(B-V)$ color index versus $V$-band magnitude diagram, which implies that the system is seen at a small or moderate inclination angle. }
    {Despite the super-Eddington X-ray luminosity during the outburst, the subsequent pattern of long-term optical and IR variability of Swift J0243.6+6124 is typical of Be/X-ray binaries.}


   \maketitle

\section{Introduction}

According to the luminosity class of the optical companion, high-mass X-ray binaries (HMXBs) are divided into supergiant X-ray binaries and Be/X-ray binaries (BeXBs; Reig \citeyear{2011Ap&SS.332....1R}).
Most of the optically identified HMXBs (or HMXB candidates) are known or suspected BeXBs (Liu et al. \citeyear{2006A&A...455.1165L}). 
The optical companion of a BeXB is a Be star, which is a nonsupergiant, fast-rotating B-type (but it may also include late O-type stars; Negueruela et al. \citeyear{2004AN....325..749N}), and luminosity class III-V star that has shown emission lines at some point in its life (Rivinius et al. \citeyear{2013A&ARv..21...69R}).
There are two different disks in Be/X-ray binaries: circumstellar disks around the Be stars, and accretion disks around the neutron stars, which temporarily appear during X-ray outbursts (Ziolkowski \citeyear{2002MmSAI..73.1038Z}; Hayasaki \& Okazaki \citeyear{2004astro.ph.12203H}).
BeXBs are classified into persistent and transient sources according to their X-ray properties (Reig \& Roche \citeyear{1999MNRAS.306..100R}). Transient BeXBs display two types of X-ray outbursts when they are active: type I (or normal) outbursts, and type II (or giant) outbursts. The peak luminosity during type I outbursts is typically $L_\mathrm{X} \leq 10^{37}$ erg s$^{-1}$, while during type II outbursts, it may reach the Eddington limit, $L_\mathrm{X} \sim 10^{38}$ erg s$^{-1}$.

\object{Swift J0243.6+6124} was detected in X-rays for the first time by the \emph{Swift}/BAT on 3 October 2017 (Kennea et al. \citeyear{2017ATel10809....1K}). It is the first Be/X-ray binary emitting at super-Eddington luminosity in our galaxy. \cite{2020A&A...640A..35R} estimated the spectral type and rotational velocity of the companion of Swift J0243.6+6124 to be O9.5Ve and $\nu \sin$ \textit{i} = 210 $\pm$ 20 km s$^{-1}$. X-ray pulsations with a period of $\sim$\,9.86 s were detected by \emph{Swift}/XRT and \emph{Fermi}/GBM (Jenke \& Wilson-Hodge \citeyear{2017ATel10812....1J}). The orbital period is 28.3 days, and the eccentricity is 0.092 (Doroshenko et al. \citeyear{2018A&A...613A..19D}).
 
There are several different reported values for the distance to this source in the literature: 2.5 $\pm$ 0.5 kpc (Bikmaev et al. \citeyear{2017ATel10968....1B}) and $\sim$\,5 kpc (Reig et al. \citeyear{2020A&A...640A..35R}), both based on optical photometric observations, a lower limit of 5 kpc set by \cite{2018Natur.562..233V}, $\sim$\,6 kpc based on two accretion torque models (Zhang et al. \citeyear{2019ApJ...879...61Z}), and $5.5^{+0.4}_{-0.3}$ kpc given in \textit{Gaia} DR3 catalog (\textit{Gaia} Collaboration et al. \citeyear{2016A&A...595A...1G,2022yCat.1355....0G}). If the distance of the source is assumed for 5 kpc, its peak luminosity is estimated as $1 \times 10^{39}$ erg s$^{-1}$ (0.1--10 keV). This exceeds the Eddington limit for the neutron star during the giant outburst.

The magnetic field strength of the neutron star in Swift J0243.6+6124 is estimated to be approximately $10^{13}$ G (Tsygankov et al. \citeyear{2018MNRAS.479L.134T}; Zhang et al. \citeyear{2019ApJ...879...61Z}), although \cite{2018A&A...613A..19D} advocated for a lower value, at the lower limit of the range $(3-9) \times10^{12}$ G. Based on the detection of electron-cyclotron resonance scattering features (CRSFs), \cite{2022arXiv220604283K} estimated a surface magnetic field of $\sim$ 1.6 $\times$ 10$^{13}$ G for Swift J0243.6+6124, which unambiguously proves the presence of multipole field components close to the surface of the neutron star. This measured surface magnetic field is the strongest of all known neutron stars with detected electron CRSFs, and it is also the strongest for all neutron star ultraluminous X-ray sources. All types of X-ray binaries have been observed to launch jets, with the exception of neutron stars that have strong magnetic fields (stronger than $10^{12}$ G), which implies that their magnetic field strength restrains jet formation (van den Eijnden et al. \citeyear{2018Natur.562..233V}). Therefore, the detection of radio emission from Swift J0243.6+6124 during the X-ray outbursts is a surprising result. However, the radio luminosity is two orders of magnitude dimmer than those seen in other accreting neutron stars with similar X-ray luminosities (van den Eijnden et al. \citeyear{2018Natur.562..233V}), which implies that the magnetic field of neutron stars still plays an important role in the power of launching jets. 
In this work, we report new optical spectroscopic observations and photometric observations. These observations witnessed the partial dissipation after the giant outburst and subsequent reformation of the Be star's circumstellar disk. 

\section{Observations}

\subsection{Optical spectroscopy} 

Optical spectroscopic observations were mainly obtained with two telescopes at two different observatories: The observations from the Xinglong Station of National Astronomical Observatories in Hebei province (China) were obtained with the spectrometer OptoMechanics Research (OMR) or BAO Faint Object Spectrograpy and Camera (BFOSC) on the 2.16 m telescope, and the observations from the Lijiang station of Yunnan Astronomical Observatory in Yunnan province (China) used the spectrometer Yunnan Faint Object Spectrograpy and Camera (YFOSC) on the 2.4 m telescope.
The OMR was equipped with a 1024 $\times$ 1024 (24 micron) pixel TK1024AB2 CCD. The OMR Grism 4 is 1200 lp $\rm mm^{-1}$, giving a nominal dispersion of 1.2 \AA\ pixel$^{-1}$, and covering the wavelength ranges 5500-6900 \AA. The BFOSC was equipped with a 2048 $\times$ 2048 (15 micron) pixel Loral Lick 3 CCD. The dispersion of BFOSC Grism 4 and 8 is 2.97 and 1.20 \AA\ pixel$^{-1}$, covering the wavelength ranges 4000--8700 and 5800--8280 \AA, respectively. The YFOSC was equipped with a 2k $\times$ 4k (13.5 micron) pixel E2V 42-90 CCD. The dispersion of YFOSC Grism 8 is 1.47 \AA\ pixel$^{-1}$, covering the wavelength ranges 4970-9830 \AA. In addition, we analyzed new optical spectroscopic observations obtained from the 1.3 m telescope of the Skinakas observatory (SKO) in Crete (Greece). The 1.3 m telescope of the SKO was equipped with a 2048 $\times$ 2048 (13.5 micron) pixel ANDOR IKON CCD and a 1302 lines $\rm mm^{-1}$ grating, giving a nominal dispersion of $\sim$\,0.8 \AA\ pixel$^{-1}$.

We used the Image Reduction and Analysis Facility (IRAF)\footnote{IRAF is distributed by NOAO, which is operated by the Association of Universities for Research in Astronomy, Inc., under cooperation with the National Science Foundation.} software package to reduce and analyze all the spectra, performing bias-subtracted correction and flat-field correction on the data, and then removing cosmic rays. A helium-argon calibration lamp was employed to obtain the pixel-wavelength relation. In order to ensure the consistency of spectral processing, all spectra were normalized to adjacent continua. We measured the equivalent width of the H$\alpha$ lines (hereafter EW(H$\alpha$) for short) for five times, each measurement with a different selection of the continuum. The final EW(H$\alpha$) is the average of the five measurements, and the error is the standard deviation. The typical error of EW(H$\alpha$) is within 5\%. The value of the error is determined by the quality of the continuum. The equivalent widths of the He I $\lambda$6678 lines (hereafter EW(He I $\lambda$6678) for short) were obtained following the same method as for EW(H$\alpha$).

The log of the spectroscopic observations is given in Table \ref{spec}. EW(He I $\lambda$6678) and EW(H$\alpha$) are plotted in the third and sixth panels of Fig.~\ref{EW}, respectively. The evolution of the H$\alpha$ line profiles is plotted in Fig.~\ref{profile}. The evolution of the $\log(V\,/\,R)$ and the peak separation of H$\alpha$ line are plotted in Fig.~\ref{V_R} and listed in Table \ref{V_R_table}. 

\subsection{Optical photometry}

Optical photometric observations were obtained from five telescopes at three different observatories: From the Xinglong station of the National Astronomical Observatories, Chinese Academy of Sciences (NAOC), observations were obtained with the Tsinghua-NAOC Telescope (TNT, 80 cm), the 60 cm telescope, and the 2.16 m telescope; from the Lijiang station of Yunnan Observatories (YNAO), the data came from the 2.4 m telescope; and from the Yaoan astronomical observation station of Purple Mountain Observatory (PMO), the data came from the Yaoan High Precision Telescope (YAHPT, 80 cm').

The TNT (80 cm) is an equatorial-mounted Cassegrain system with a focal ratio of f/10, made by AstroOptik, funded by Tsinghua University in 2004 and jointly operated with NAOC, which is equipped with a PI VersArray 1300B LN 1340 $\times$ 1300 thin, back-illuminated CCD with a 20 $\mu$m pixel$^{-1}$ size \citep{2008ApJ...675..626W,2012RAA....12.1585H}.  In this configuration, the plate scale is 0.52" pixel$^{-1}$ and gives a field of view of $11.4 \times 11.1$ $\rm arcmin^{2}$.
The 60 cm telescope is an equatorial-mounted system with a focal ratio of f/4.23, which is equipped with the Andor DU934P-BEX2-DD 1024 $\times$ 1024 CCD and provides a field of view of 18 $\times$ 18 $\rm arcmin^{2}$.
The 2.4 m telescope is an altazimuth-mounted Cassegrain system with a focal ratio of f/8,  which is equipped with an E2V CCD42-90  2k $\times$ 2k thin, back-illuminated, deep-depletion CCD with a 13.5 $\mu$m pixel$^{-1}$ size. In this configuration, the plate scale is 0.28" pixel$^{-1}$ and gives a field of view of 9.6 $\times$ 9.6 $\rm arcmin^{2}$.
The YAHPT (80 cm') is an altazimuth-mounted, RC optical system with a focal ratio of f/10, made by Astro Systeme Austria, which is equipped with a PIXIS 2048B back-illuminated CCD with a 13.5 $\mu$m pixel$^{-1}$ size.  In this configuration, the plate scale is 0.347" pixel$^{-1}$, providing a field of view of 11.8 $\times$ 11.8 $\rm arcmin^{2}$.
The 2.16 m telescope is an equatorial-mounted, RC optical system with a focal ratio of f/9, made by NAOC, CAS Nanjing Astronomical Instruments Co., LTD (NAIRC), and the Institute of Automation of the Chinese Academy of Sciences (CASIA), which is equipped with an Andor-DZ936-BEX2-DD 2048 $\times$ 2048 CCD with a 13.5 $\mu$m pixel$^{-1}$ size.  In this configuration, the plate scale is 0.274" pixel$^{-1}$, providing a field of view of 9.36 $\times$ 9.36 $\rm arcmin^{2}$.

In all five telescopes, Swift J0243.6+6124 was observed through the standard Johnson-Cousins $B$, $V$, $R$, and $I$ filters. The photometric data reduction was performed using standard routines and aperture photometry packages (some from the zphot package) in IRAF, including bias subtraction and flat-field correction. In order to derive the variation in the optical brightness, we selected the reference star Gaia 465628266540345216 ($\alpha$: 02 43 38.23, and $\delta$: +61 26 40.7, J2000) \cite[according to ][ the average magnitudes of the reference star are B = 13.67 $\pm$ 0.01, V = 13.02 $\pm$ 0.01, R = 12.65 $\pm$ 0.01, and I = 12.25 $\pm$ 0.02]{2020A&A...640A..35R} in the field of view of Swift J0243.6+6124 to derive its differential magnitudes. The photometric magnitudes are given in Table \ref{phot}.

To study the long-term optical variability of the source, we used the public optical photometric data from the \emph{ASAS–SN}\footnote{https://asas-sn.osu.edu/variables/7306192e-fb93-58a1-98a8-1809\\e318a711} Variable Stars Database (Shappee et al. \citeyear{2014ApJ...788...48S}; Jayasinghe et al. \citeyear{2019MNRAS.486.1907J}). There is a slightly fainter star at 6.2 arcsec from Swift J0243.6+6124 within the full width at half maximum (FWHM). This star is resolved in our photometry. However, the pixel scale and the FWHM in ASAS–SN are 8 arcsec and $\sim$2 pixels, and hence in these images, the neighboring star contributes to the measured flux from Swift J0243.6+6124. The calibrated $V$-band magnitude of the fainter close star is $V$ = 14.52 $\pm$ 0.01 mag (Reig et al. \citeyear{2020A&A...640A..35R}). We removed the brightness of the neighboring star from the total observed flux. The applied corrections are in the range $\Delta$V = 0.20$^{+0.03}_{-0.04}$ mag. We also made use of the public optical photometric data from the international database of the American Association of Variable Star Observers (\emph{AAVSO}\footnote{https://app.aavso.org/webobs/results/?star=000-BML-322\&num\_\\results=200}). Finally, we also included the optical photometric data from \cite{2020A&A...640A..35R}. 

The Johnson $V$-band light curve is plotted in the fourth panel of Fig.~\ref{EW}. A detailed view of the 2017 outburst, including pre- and post-outburst observations, is shown in Fig.~\ref{V_IR}. The evolution of the $(B-V)$ color index is plotted in the seventh panel of Fig.~\ref{EW}, and the variation of the $(B-V)$ color index versus $V$-band magnitude is plotted in Fig.~\ref{B-V_V}, where only the data from the 80 cm telescope, the 2.4 m telescope, and the 2.16 m telescope are shown.

\subsection{\emph{NEOWISE} photometry}

We made use of the light curves in the W1 (3.4 $\mu$m) and W2 (4.6 $\mu$m) bands provided by the \emph{NEOWISE} (Mainzer et al. \citeyear{2011ApJ...731...53M}) project  through the IRSA viewer\footnote{https://irsa.ipac.caltech.edu/irsaviewer}. We plot them in the fifth panel of Fig.~\ref{EW} and in the bottom panel of Fig.~\ref{V_IR}.

\subsection{X-Ray observations}

The Burst Alert Telescope (BAT)\footnote{https://swift.gsfc.nasa.gov/results/transients/weak/SwiftJ0243.6p61\\24/} on board \emph{Swift} (Krimm et al. \citeyear{2013ApJS..209...14K}), \emph{MAXI}\footnote{http://maxi.riken.jp/star\_data/J0243+614/J0243+614.html} , and the  Gamma-ray Burst Monitor (GBM)\footnote{https://gammaray.nsstc.nasa.gov/gbm/science/pulsars/lightcurves/\\swiftj0243.html} on board \emph{Fermi} (Meegan et al. \citeyear{2009ApJ...702..791M}) have been monitoring Swift J0243.6+6124 in the hard X-ray energy band (15–50 keV with BAT, 2--20 keV with \emph{MAXI,} and 12--50 keV with GBM) since October 2017. One type II X-ray outburst and several type I outbursts were detected between October 2017 and January 2019. The X-ray band light curves from BAT (15–50 keV) and \emph{MAXI} (2--20 keV) are plotted in the first panel of Fig.~\ref{EW}. The spin-frequency history measured with GBM is plotted in the second panel of Fig.~\ref{EW}.

\section{Results}

\begin{figure*}
   \centering
   \includegraphics[bb=-200 300 1500 2700,width=15cm]{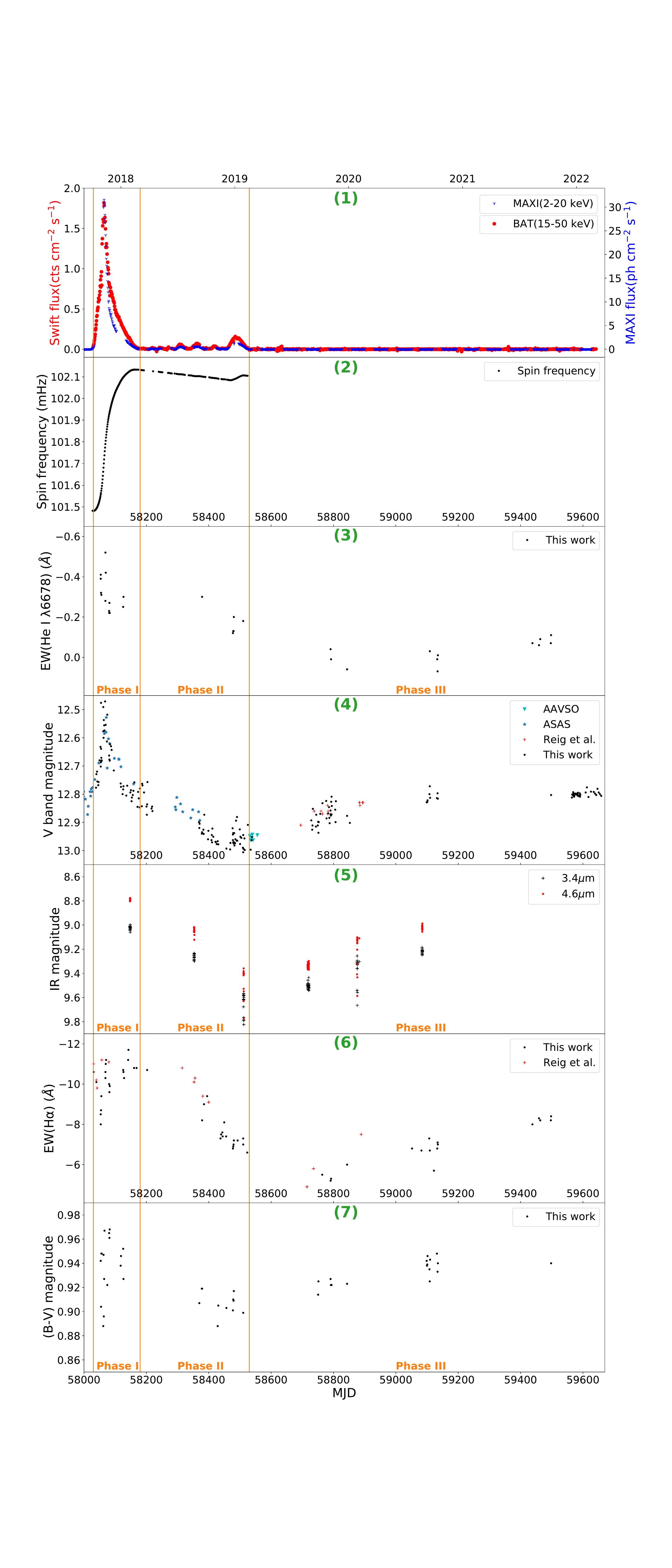}
   \caption{Long-term variations of Swift J0243.6+6124 from 2017 to 2022. First panel: X-ray band light curves. The data from \emph{Swift}/BAT (15--50 keV) and \emph{MAXI} (2--20 keV) are labeled with red points and blue downward triangles, respectively. The first day of each year is also marked at the top of the first panel in the format YYYY.
   Second panel: Barycentered and orbit-corrected spin-frequency history measured with \emph{Fermi}/GBM. 
   Third panel: Equivalent widths of He I $\lambda$6678 lines.
   Fourth panel: Long-term light curves of the optical $V$ band from \emph{AAVSO}, ASAS-SN, \cite{2020A&A...640A..35R}, and this work. They are separately labeled with light green downward triangles, blue stars, red plus signs, and black points.  
   Fifth panel: Long-term light curves of \emph{NEOWISE} infrared magnitudes at 3.4 $\mu$m and 4.6 $\mu$m. They are labeled with black plus signs and red points, respectively. 
   Sixth panel: Equivalent widths of H$\alpha$ lines from this work and \cite{2020A&A...640A..35R}. They are labeled with black points and red plus signs, respectively. 
   Seventh panel: Evolution of the $(B-V)$ color index.
   The vertical lines separate the three phases into which the observations are divided.
   }
    \label{EW}%
    \end{figure*}

Figure~\ref{EW} shows the X-ray, optical, and IR long-term variability of  Swift J0243.6+6124. The observations cover the October 2017 X-ray outburst and the changes experienced by the source during the following four-year period. After the X-ray outbursts, the optical brightness of the source decreased and reached a minimum on $\sim$ MJD 58530. Since then, it has been recovering. There is a clear correlation between the optical and IR flux on long timescales (weeks or months). When the source is active in the X-ray, the optical and IR also correlate with the overall X-ray flux in the sense that the source is bright in the optical and IR at the time of the outbursts. The strength of the H$\alpha$ line also follows the same general trend, although the minimum after the X-ray outbursts appears to be delayed by $\sim$\,100--200 days with respect to the optical or IR continuum flux.

After examining the long-term light curves and spectral evolution, we divided the observations into three different epochs or phases (Fig.~\ref{EW}). These phases reflect significant changes in the properties of the data. Each phase is characterized by a different pattern of X-ray, optical, or IR variability. Phase I corresponds to the giant 2017 X-ray outburst; during phase II, the source experiences a gradual fading of its brightness and a weakening of the spectral line parameters; and in phase III, the long-term trend is reversed and the source exhibits a gradual increase in the brightness and strength of the spectral lines, most significantly, in the H$\alpha$ line. 

Phase I (MJD 58030--58180) covers the giant X-ray outburst. The X-ray luminosity changes by about two orders of magnitude ($10^{37}-10^{39}$ erg s$^{-1}$;  Doroshenko et al. \citeyear{2020MNRAS.491.1857D}) and the $V$ band by 0.4 magnitudes. EW(H$\alpha$) and EW(He I $\lambda$6678) display erratic variability. The $(B-V)$ color displays the largest variation in the entire period of the observations, with a change of about 0.1 magnitude in 25 days.

In Phase II (MJD 58180--58530), the X-ray variability is characterized by regular type I outbursts with changes in luminosity of about one order of magnitude between $10^{36}-10^{37}$ erg s$^{-1}$. The brightness in the $V$ band decreases by 0.1 magnitudes and in the near-infrared by 0.6 magnitudes. EW(H$\alpha$) decreases from a maximum of $-11$ \AA\ to a minimum of $-5$ \AA. EW(He I $\lambda$6678) presents large scatter, but it also decreases on average. The dispersion in EW(He I $\lambda$6678) measurements is most likely due to the low signal-to-noise ratio (S/N). There are no suitable B-band observations during the first half of this phase; thus we do not have $(B-V)$ data. During the second half, the $(B-V)$ color is lower on average (i.e., bluer emission) than during phase I.

In Phase III (MJD 58530--), the source is no longer detected in X-ray, while all optical and infrared indicators increase gradually. At the end of this phase, EW(H$\alpha$) and the $V$-band magnitude recover to almost pre-outburst values. The overall optical emission becomes redder as the system evolves in this phase.

\subsection{ H$\alpha$ line profiles}

Figure~\ref{profile} shows the profiles of the H$\alpha$ line. Although we are limited by the low spectral resolution, some general variability trends are visible. The H$\alpha$ emission lines present asymmetrically blue-dominated profiles (V>R) between October 2017 and September 2018 (phase I--phase II). The red peak is only noticeable in the earliest spectrum on 27 October  2017. The central rest wavelength (marked by the vertical dashed line in Fig.~\ref{profile}) lies systematically to the right of the peak, confirming the asymmetry. At the end of 2018 and during 2019 (phase II--phase III), the line flux gradually shifts toward the red. Approximately symmetric single-peak profiles are observed in the 2020-2021 spectra (phase III). At the end of 2021, the profiles are again blue-dominated.

The He I line also exhibits a variable asymmetric double-peak profile in phase and with intensity variations similar to those of the H$\alpha$ line. Because the He I line forms much closer to the Be star than the H$\alpha$ line, the He I V\,/\,R variability implies that the density perturbation affects the inner parts of the disk as well as the outer parts where the H$\alpha$ line is formed. Unfortunately, the low S/N and low spectral resolution prevented a detailed analysis of this line.

\section{Discussion}

\begin{figure*}
   \centering
   \includegraphics[bb=0 0 1250 800,width=\hsize]{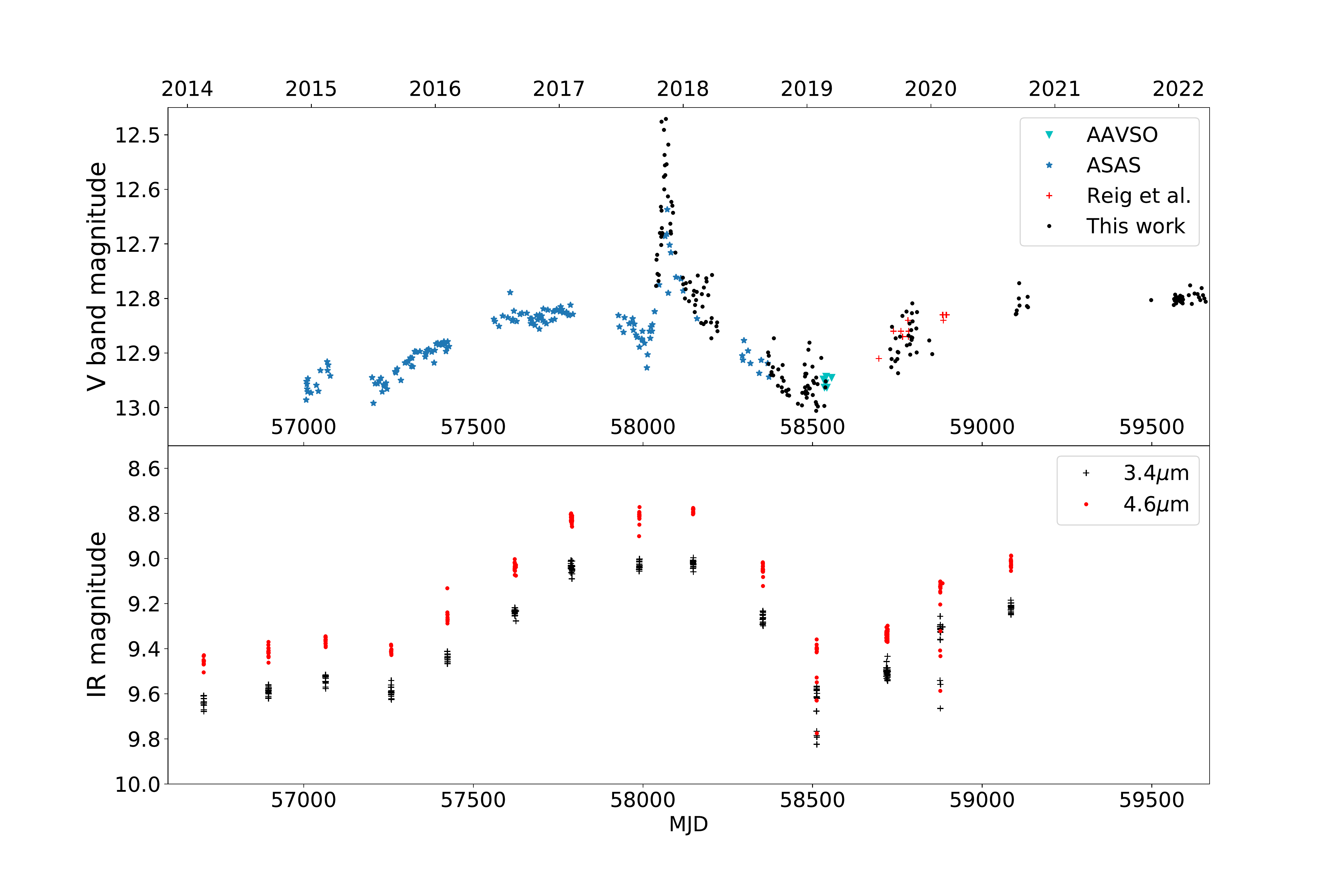}
      \caption{Multiwavelength light curves of Swift J0243.6+6124. Top panel: Long-term light curves of the optical $V$ band from MJD 56600 to 59670. 
      Bottom panel: Simultaneous light curves of \emph{NEOWISE} infrared magnitudes at 3.4 $\mu$m and 4.6 $\mu$m.
              }
         \label{V_IR}
   \end{figure*}

We have been monitoring the optical counterpart of the Be/X-ray binary Swift J0243.6+6124 since its discovery in October 2017 and obtained optical spectra on a regular basis. Using data from both the archives and the literature, we are able to study the long-term variability over at least four years after the outburst. We attribute this long-term variability to the evolution of the Be star's circumstellar disk.

The variability patterns of the X-ray, optical, and IR parameters allowed us to divide the observations into three different phases. Phase I covers the rise and decay of the X-ray outburst. The largest and fastest variability timescales are seen during this phase. This violent event must have affected the entire structure of the circumstellar disk. In particular, the H$\alpha$ line displayed several sudden changes of 10\%--20\% in its strength on timescales of $\text{about one}$\, month during this phase (see the sixth panel of Fig.~\ref{EW}). These fast changes could be attributed to an inhomogeneous warped disk (Reig et al. \citeyear{2020A&A...640A..35R}). Alternatively, because the fast changes of EW(H$\alpha$) appear to be modulated with the orbital period of the system ($P_{\rm orb}=28.3$ days), reprocessed emission might be at their origin. The high spin-up rate (as seen in the second panel of Fig.~\ref{EW}) suggests that a short-lived accretion disk also formed around the neutron star. 

Phase II corresponds to a low optical state in which not only the optical and IR continuum flux gradually decreased, but the H$\alpha$ emission line also became weaker. Type II outbursts involve the accretion of a large amount of material from the equatorial disk and almost always lead to the dissipation of the entire or part of the disk \citep{2016A&A...590A.122R}. We identify this phase with the dissipation of the disk.
The evolution of the X-ray and optical emissions provides some clues about how this dissipation took place. The presence of type I outbursts indicates that the neutron star accretes material from the disk in subsequent periastron passages, while at the same time, the decrease in the $(B-V)$ color indicates that the emission becomes bluer, as expected from a smaller or more compact disk. These results can be understood if the outer parts of the disk cease to be bound, that is, are expelled, while the inner parts of the disk collapse toward the star once the disk formation mechanism stops. The higher X-ray intensity of the last type I outburst in this phase may be due to the fact that the neutron star encountered a higher-density part of this distorted disk. This phase ends when the magnitude and colors reach a minimum. The system did not lose the disk entirely, as the H$\alpha$ line remained with an emission profile and the minimum EW(H$\alpha$) was still $\sim$\,$-5$ \AA. Even the He I $\lambda$6678 did not revert into an absorption profile. The equivalent width of this line remained $\sim$\,0 \AA\ during 2020 and 2021, indicating that the line was filled with emission. 

Phase III represents a phase in which the disk grew again. The optical and IR flux increased, as did the strength of the H$\alpha$ line. The overall emission became redder (i.e., $(B-V)$ increased). The latest observations show that EW(H$\alpha$) and EW(He I $\lambda$6678) approached pre-outburst values. After an initial brightening period that lasted for about a year, the $V$-band magnitude stabilized at a level of 12.8${^m}$, at which it has remained since September 2020.

Photometric variability with a characteristic period of $\sim$\,1250 days (MJD 40000--41250) and an amplitude of $\sim$\,0.15 magnitudes in the B band was reported by \cite{2017ATel10989....1N} based on archival data from the Asiago Observatory taken between 1967 and 1976. This timescale is similar to the one during MJD 57250 to 58530 (see Fig.~\ref{V_IR}). When the $V$-band flare is ignored, the underlying trend of the $V$-band observations correlates very well with the IR-band observations. Given the strength of this correlation, the lack of a similar flare in the IR light curve can be attributed to the low cadence of the IR observations. The disk starts to grow on $\sim$ MJD 57250, reaching a maximum size on $\sim$ MJD 58000 when it begins to decline, and reaching a minimum on $\sim$ MJD 58530. 
The variation in the $V$/IR-band can be interpreted in terms of the evolution of the Be star's disk. This long-term smooth change in brightness is due to the formation or growth of the disk and its subsequent dissipation. 
We estimate that the overall timescale in phase III for the formation and dissipation of the circumstellar disk is about 1500 days. The dissipation phase is significantly faster, about 300 days, than the formation phase of about 1200 days.
   
Although the evolution of the optical and IR parameters is affected by observational gaps, because the position of the source is too close to the Sun, Fig.~\ref{EW} shows a delay between the minima of the optical and IR continuum (which marks the shift from phase II to phase III) and the minima of EW(H$\alpha$). This delay may be understood by invoking the different sites of the continuum and discrete emission in the equatorial disk of a Be star. According to \cite{2011IAUS..272..325C}, the disk $V$ band is typically formed very close to the star, within about 2\,$R_{\rm star}$. In contrast, the H$\alpha$ emission line is formed at larger radii (Slettebak et al. \citeyear{1992ApJS...81..335S}).  If this interpretation is correct, then the minimum flux detected in the continuum first would imply that the dissipation of the disk began from the inner parts. The type I regular X-ray outbursts during phase II indicate that the accretion of the neutron star persisted for about one year after the main outburst. In terms of the strength of the H$\alpha$ line, the giant X-ray outburst took place when EW(H$\alpha$) $\sim$\,$-11$ \AA. The increase rate of EW(H$\alpha$) during phase III implies that the source will reach this value again on $\sim$\,MJD 60250. If we take this value ($-11$ \AA) as the triggering value of the giant outburst, then we should expect another large event by the end of 2023.

The He I $\lambda$6678 follows the same long-term trend as the H$\alpha$ line: fast and large amplitude changes during phase I, a weakening during phase II, and a slow recovery during phase III. The latest spectrum in October 2021, in which a small peak started to develop and EW(He I $\lambda$6678) < $-0.1$ \AA, marks the formation of a new emission line. We note that the giant X-ray outburst occurred when EW(He I $\lambda$6678) $\sim$\,$-(0.2-0.5)$ \AA.

\subsection{ H$\alpha$ line profile variability and V\,/\,R ratio}

\begin{figure}
   \centering
   \includegraphics[bb=50 100 700 1300,width=\hsize]{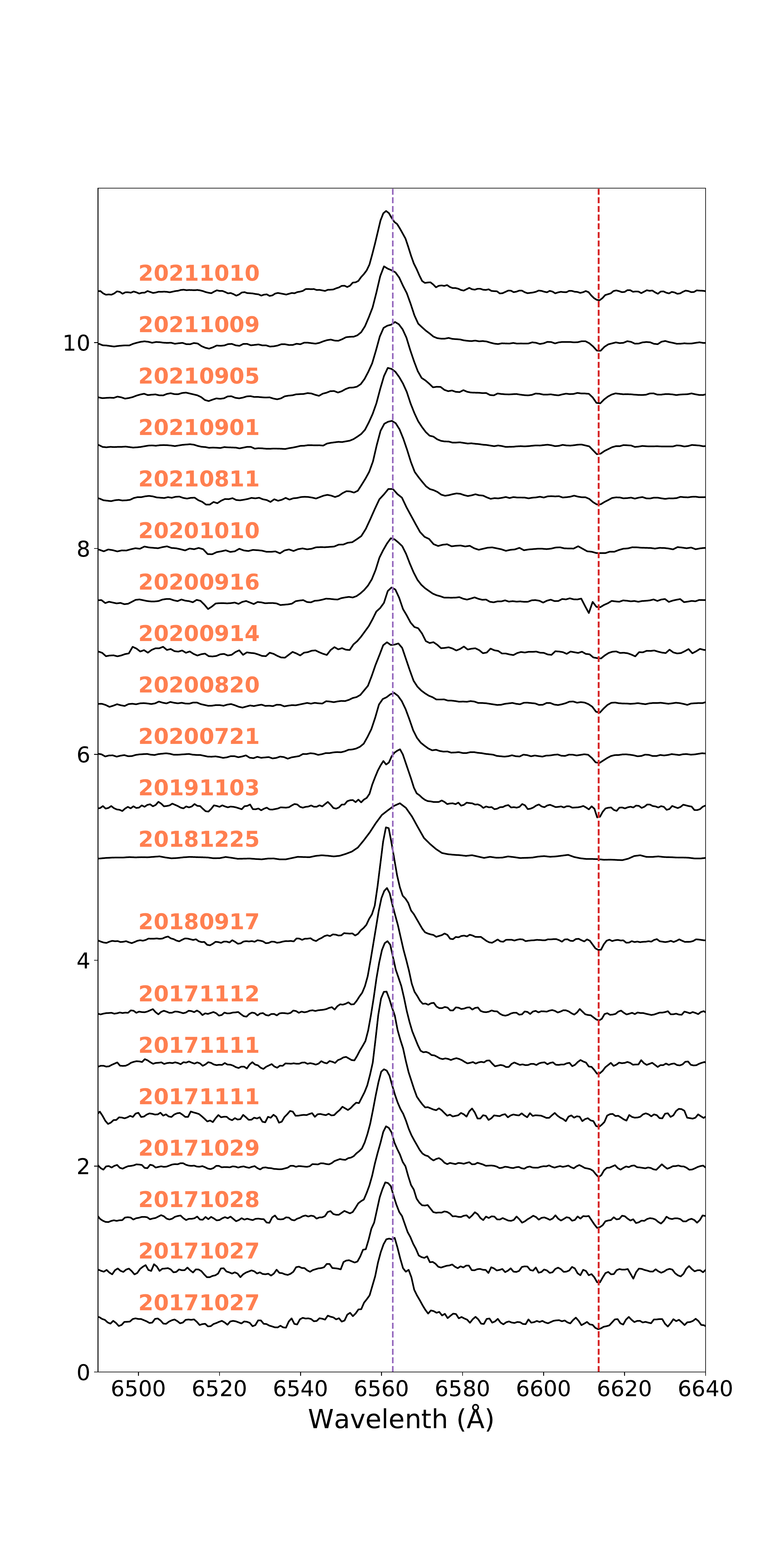}
      \caption{Evolution of H$\alpha$ line profiles. The vertical lines mark the rest wavelength of the H$\alpha$ lines and diffuse interstellar bands (DIB) 6613 \AA. All spectra have been normalized with adjacent continua.}
         \label{profile}
   \end{figure}
   
\begin{figure}
   \centering
   \includegraphics[bb=0 0 650 400,width=\hsize]{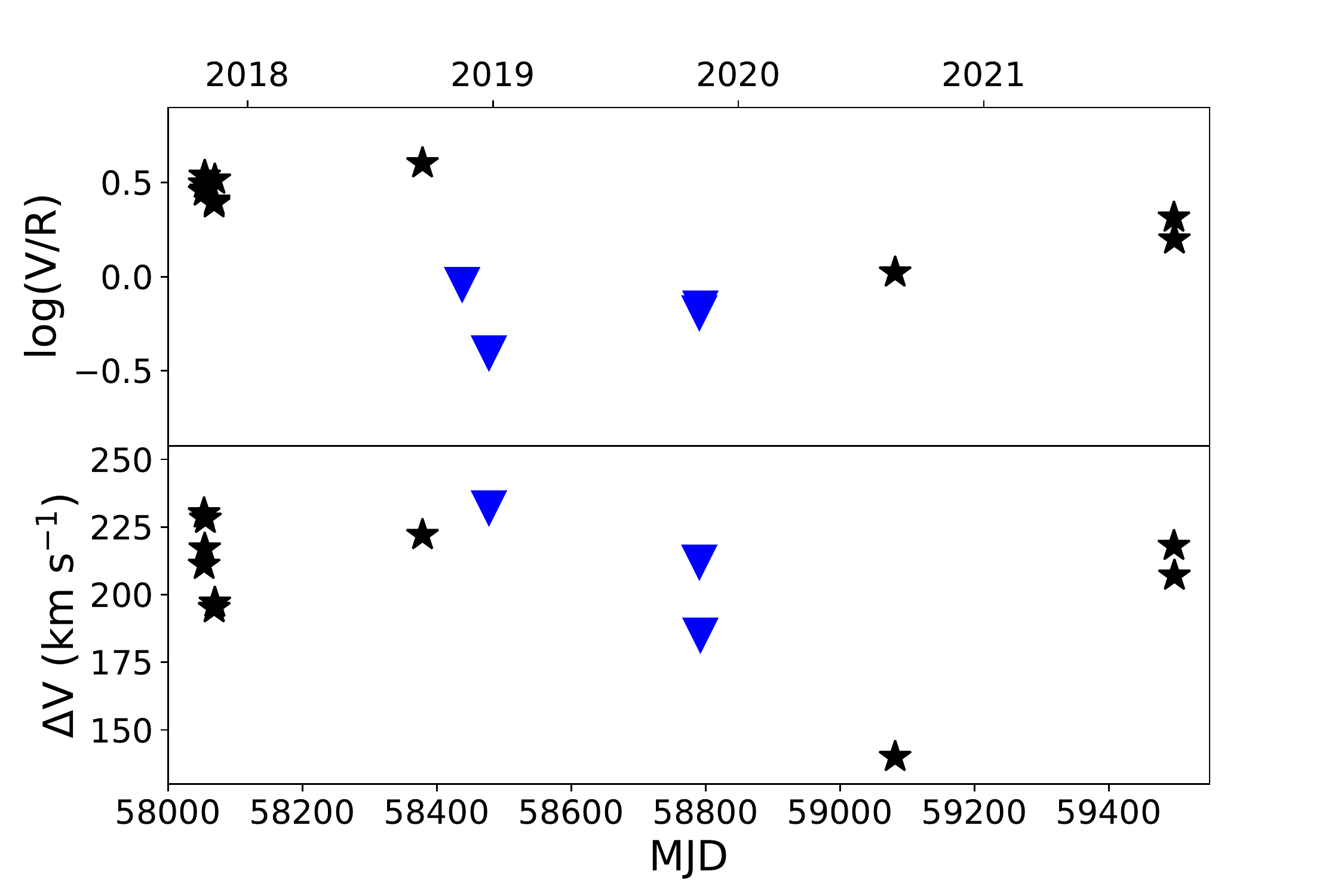}
      \caption{Variations in spectral features of H$\alpha$ line. Top panel: Evolution of $\log(V\,/\,R)$. 
      Bottom panel: Peak separation of H$\alpha$ line. 
      Stars represent the spectral profiles with V>R, and triangles represent the spectral profiles with V<R.
              }
         \label{V_R}
   \end{figure}

The V\,/\,R variability is defined as the intensity variations of the two peaks (known as violet and red peaks) in the split profile of a spectral line. In many Be stars, if they are monitored over a long enough period of time, these variations are quasi-periodic (Okazaki \citeyear{1997A&A...318..548O}).
We define the V\,/\,R ratio of the H$\alpha$ line as V\,/\,R = (I(V) $-$ I$_\mathrm{c}$) / (I(R) $-$ I$_\mathrm{c}$), where I(V), I(R), and I$_\mathrm{c}$ are the intensities of the violet peak, red peak, and continuum, respectively. 
We also measured the separation of the violet and red peaks by fitting two Gaussian functions to the spectral line profile. When the disk velocity is assumed to be Keplerian, the peak separation gives a measure of the velocity field.
There is no obvious trend in the peak separation between different spectral line profiles, which is mainly distributed around 175--250 km s$^{-1}$.
The V\,/\,R ratios and the peak separation of the H$\alpha$ line are listed in Table \ref{V_R_table} and plotted in Fig.~\ref{V_R}.

The V\,/\,R variability has been associated with density perturbations in the disk (Hanuschik et al. \citeyear{1995A&A...300..163H}). When this density perturbation moves around inside the disk, the profile changes. 
We observe a blue-dominated profile (V>R) in 2017 that turned into an almost single peak profile (V$\sim$R) in 2018 (see Fig.~4 of Reig et al. \citeyear{2020A&A...640A..35R}), and a red-dominated profile (V<R) from the end of 2018 to 2019.
The spectra in 2021 return to blue-dominated profiles (V>R). Thus, we may have covered an entire V\,/\,R cycle. The V\,/\,R quasi-period would be $\text{about four}$\, years, which is normal for BeXBs (Mennickent et al. \citeyear{1997A&A...326.1167M}).

In principle, the question of whether the motion of the perturbation occurs in the same sense (prograde rotation) or opposite sense (retrograde rotation) to the stellar rotation can be determined from the observations. 
\cite{1994A&A...288..558T} realized that a prograde rotation implies that (I) a V>R phase, (II) a shell absorption profile, (III) a V<R phase, and (IV) a weak central absorption profile will appear in order. A retrograde rotation would give rise to the reversed sequence: (IV)$\to$(III)$\to$(II)$\to$(I). Because of the small disk inclination (Reig et al. \citeyear{2020A&A...640A..35R}), we cannot distinguish a prograde or retrograde rotation in the characteristic line shapes. These characteristic line shapes can translate into noticeable photometric variations, however. 
According to \cite{1997A&A...326.1167M}, we can expect a minimum brightness when V=R prior to the V<R (V>R) phase if the motion is prograde (retrograde).
In Swift J0243.6+6124, the minimum brightness in the photometric $V$ band occurred during the V=R phase before the V<R phase began, $\sim$\,MJD 58450, confirming the prograde nature of the precession inside the disk.

\subsection{Variation in $(B-V)$ color index and  inclination of the Be star's disk}

\begin{figure}
   \centering
   \includegraphics[bb=0 0 600 400,width=\hsize]{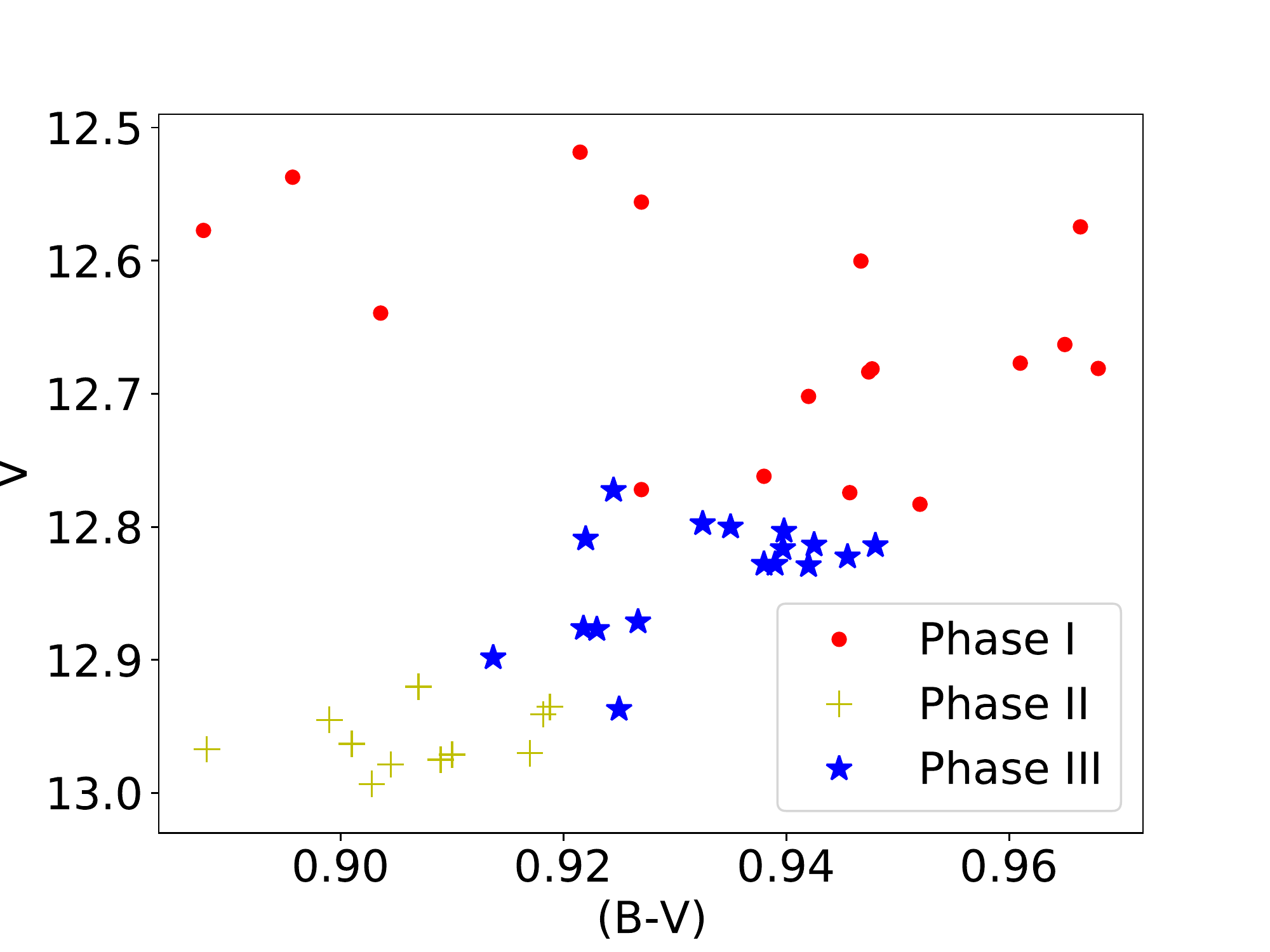}
      \caption{Variation in $(B-V)$ color index versus $V$-band magnitude.
              }
         \label{B-V_V}
   \end{figure}

Figure~\ref{B-V_V} shows the $(B-V)$ color index as a function of the $V$ magnitude. We mark the different variability phases defined above with different colors. It has been noted that this kind of plot can be used to constrain the inclination angle of the system \citep{1983HvaOB...7...55H}.
Systems that show a positive correlation, that is, as the disk forms (or equivalently, as EW(H$\alpha$) increases), the optical intensity increases and the emission becomes redder (i.e., $(B-V)$ increases), are thought to be seen at small or moderate inclination angles, while systems that show a negative correlation, in which the optical intensity decreases even though the disk is growing (EW(H$\alpha$) and $(B-V)$ increase), are associated with large inclination angles.

\cite{1983HvaOB...7...55H} introduced the concept of a pseudophotosphere to explain this effect. At large inclination angles (for equator-on stars), the inner parts of the Be envelope partly block the stellar photosphere, and thus the optical brightness decreases. Meanwhile, the overall emission becomes redder because the contribution of the disk increases. At small or intermediate inclination, as the disk grows, an overall (star plus disk) increase in brightness is expected.

Figure~\ref{B-V_V} suggests that Swift J0243.6+6124 is viewed at small or intermediate angles. This result is consistent with the 30$^{\circ}$ angle estimated by \cite{2020A&A...640A..35R} from the emission-line profile.

\section{Conclusions}

We have conducted spectroscopic and photometric observations of Swift J0243.6+6124 to study the changes in the structure of the circumstellar disk of the Be star and the material transfer between the Be star's circumstellar disk and the neutron star during 2017--2022. This period covers a giant X-ray outburst (type II) and several orbit-modulated outbursts (type I). 
We divided our data into three phases based on the intensity of the X-ray, optical, and infrared emission. 
Phase I covers the 2017 X-ray outburst. In this phase, the source reaches the largest equivalent width of the H$\alpha$ and He I $\lambda$6678 lines and the brightest X-ray, optical, and infrared intensity, characterized by outbursts or flares.
In phase II, the source displays a long-term decrease in the optical and infrared intensities. The X-ray band is characterized by the occurrence of several minor orbit-modulated outbursts.
In phase III, the source exhibits a long-term brightening of the optical and infrared magnitudes, and the equivalent width of the H$\alpha$ and He I $\lambda$6678 lines also increases. During this phase, no X-ray activity is observed. 
We focused on the correlated optical, IR, and X-ray long-term variability. During the period of our observations following the giant X-ray outburst, the optical and IR continuum flux and the strength of the H$\alpha$ line first decreased and then increased. We interpret this long-term variability in terms of the dissipation and reformation of the Be star's circumstellar disk. 
Although the 2017 giant X-ray outburst is the most luminous ever recorded in a BeXB, it does not lead to the complete loss of the Be star's disk. 
We estimate that at the rate at which the disk is reforming, the equivalent width of the H$\alpha$ line will reach pre-outburst values in about 1--1.5 years. 

\begin{acknowledgements}
      We acknowledge the support of the staff of the Xinglong 2.16 m telescope, the Xinglong 80 cm telescope, and the Xinglong 60 cm telescope. This work was partially supported by the Open Project Program of the CAS Key Laboratory of Optical Astronomy, National Astronomical Observatories, Chinese Academy of Sciences.
      
      We acknowledge the support of the staff of the Lijiang 2.4 m telescope. Funding for the telescope has been provided by CAS and the People's Government of Yunnan Province.
      
      Skinakas Observatory is run by the University of Crete and the Foundation for Research and Technology-Hellas.
      
      This research has made use of data provided by the Yaoan High Precision Telescope.
      
      We acknowledge with thanks the variable star observations from the \emph{AAVSO} International Database contributed by observers worldwide and used in this research.
      
      \emph{Swift}/BAT transient monitor results provided by the \emph{Swift}/BAT team. \emph{Fermi}/GBM results provided by the Fermi Science Support Center.
      
      This publication makes use of data products from \emph{NEOWISE}, which is a project of the Jet Propulsion Laboratory/California Institute of Technology, funded by the Planetary Science Division of the National Aeronautics and Space Administration.
      
      This work is supported by the National Key R\&D Program of China (2021YFA0718500) and the National Natural Science Foundation of China (Grants No. U2031205, 11733009).
      
      X. Wang is supported by the National Science Foundation of China (NSFC grants 12033003 and 11633002), the Scholar Program of Beijing Academy of Science and Technology (DZ:BS202002), and the Tencent Xplorer Prize.
\end{acknowledgements}


\begin{appendix}

\section{Table of spectroscopic observations}

\begin{table*}
\caption{Spectroscopic observations of Swift J0243.6+6124 between 2017 and 2021.}
\label{spec}
 \centering
\begin{tabular}{cclccc}
\hline\hline
Date & MJD & Telescope/ & Wavelength Range & EW(H$\alpha$) & EW(He I $\lambda$6678) \\
(DD-MM-YYYY) & & Instrument & (\AA) & (\AA) & (\AA) \\
\hline
 04-10-2017 & 58030.8342 & 2.16 m/BFOSC & 4000-8700 & -10.6 $\pm$ 0.2 & ...              \\
 13-10-2017 & 58039.6745 & 2.16 m/BFOSC & 4000-8700 & -10.1 $\pm$ 0.1 & ...              \\
 27-10-2017 & 58053.6744 & 2.16 m/OMR   & 5500-6900 & -8.5  $\pm$ 0.1 & -0.39 $\pm$ 0.08 \\
 27-10-2017 & 58053.8025 & 2.16 m/OMR   & 5500-6900 & -8.0  $\pm$ 0.4 & -0.41 $\pm$ 0.11 \\
 28-10-2017 & 58054.6191 & 2.16 m/OMR   & 5500-6900 & -8.7  $\pm$ 0.4 & -0.32 $\pm$ 0.12 \\
 29-10-2017 & 58055.6853 & 2.16 m/OMR   & 5500-6900 & -9.4  $\pm$ 0.3 & -0.31 $\pm$ 0.16 \\
 11-11-2017 & 58068.6412 & 2.16 m/OMR   & 5500-6900 & -10.3 $\pm$ 0.6 & -0.28 $\pm$ 0.23 \\
 11-11-2017 & 58068.7547 & 2.16 m/OMR   & 5500-6900 & -10.6 $\pm$ 0.2 & -0.52 $\pm$ 0.12 \\
 12-11-2017 & 58069.7245 & 2.16 m/OMR   & 5500-6900 & -11.0 $\pm$ 0.5 & -0.42 $\pm$ 0.09 \\
 13-11-2017 & 58070.7783 & 2.16 m/BFOSC & 4000-8700 & -11.2 $\pm$ 0.2 & ...              \\
 23-11-2017 & 58080.5667 & 2.4 m/YFOSC  & 4970-9830 & -10.0 $\pm$ 0.2 & -0.23 $\pm$ 0.06 \\
 23-11-2017 & 58080.6792 & 2.4 m/YFOSC  & 4970-9830 & -10.0 $\pm$ 0.2 & -0.22 $\pm$ 0.06 \\
 24-11-2017 & 58081.5326 & 2.4 m/YFOSC  & 4970-9830 & -9.6  $\pm$ 0.2 & -0.27 $\pm$ 0.06 \\
 25-11-2017 & 58082.5549 & 2.4 m/YFOSC  & 4970-9830 & -9.9  $\pm$ 0.1 & -0.22 $\pm$ 0.03 \\
 07-01-2018 & 58125.6306 & 2.4 m/YFOSC  & 4970-9830 & -10.7 $\pm$ 0.1 & -0.25 $\pm$ 0.09 \\
 08-01-2018 & 58126.5597 & 2.4 m/YFOSC  & 3700-7630 & -10.6 $\pm$ 0.2 & -0.30 $\pm$ 0.06 \\
 10-01-2018 & 58128.5208 & 2.16 m/BFOSC & 4000-8700 & -10.3 $\pm$ 0.4 & ...              \\
 23-01-2018 & 58141.5189 & 2.16 m/BFOSC & 4000-8700 & -11.2 $\pm$ 0.1 & ...              \\
 24-01-2018 & 58142.5440 & 2.16 m/BFOSC & 4000-8700 & -11.7 $\pm$ 0.1 & ...              \\
 12-02-2018 & 58160.4380 & 2.16 m/BFOSC & 4000-8700 & -10.8 $\pm$ 0.1 & ...              \\
 19-02-2018 & 58168.4470 & 2.16 m/BFOSC & 4000-8700 & -10.8 $\pm$ 0.2 & ...              \\
 25-03-2018 & 58202.4619 & 2.16 m/BFOSC & 4000-8700 & -10.7 $\pm$ 0.3 & ...              \\
 17-09-2018 & 58378.7433 & 2.16 m/OMR   & 5500-6900 & -8.2  $\pm$ 0.5 & -0.30 $\pm$ 0.03 \\
 23-09-2018 & 58384.7466 & 2.16 m/BFOSC & 4000-8700 & -9.0  $\pm$ 0.1 & ...              \\
 03-10-2018 & 58394.8377 & 2.16 m/BFOSC & 4000-8700 & -9.4  $\pm$ 0.1 & ...              \\
 15-11-2018 & 58437.6707 & 2.16 m/BFOSC & 4000-8700 & -7.3  $\pm$ 0.6 & ...              \\
 17-11-2018 & 58439.6922 & 2.16 m/BFOSC & 4000-8700 & -7.5  $\pm$ 0.2 & ...              \\
 21-11-2018 & 58443.5873 & 2.16 m/BFOSC & 4000-8700 & -7.6  $\pm$ 0.0 & ...              \\
 22-11-2018 & 58444.6232 & 2.16 m/BFOSC & 4000-8700 & -7.4  $\pm$ 0.2 & ...              \\
 27-11-2018 & 58449.6591 & 2.16 m/BFOSC & 4000-8700 & -8.1  $\pm$ 0.1 & ...              \\
 04-12-2018 & 58455.7521 & 2.16 m/BFOSC & 4000-8700 & -7.4  $\pm$ 0.7 & ...              \\
 25-12-2018 & 58477.5579 & 2.4 m/YFOSC  & 4970-9830 & -6.8  $\pm$ 0.1 & -0.12 $\pm$ 0.06 \\
 26-12-2018 & 58478.5499 & 2.4 m/YFOSC  & 4970-9830 & -6.9  $\pm$ 0.2 & -0.13 $\pm$ 0.06 \\
 27-12-2018 & 58479.5959 & 2.4 m/YFOSC  & 4970-9830 & -7.0  $\pm$ 0.2 & -0.13 $\pm$ 0.05 \\
 28-12-2018 & 58480.5463 & 2.4 m/YFOSC  & 4970-9830 & -7.2  $\pm$ 0.2 & -0.20 $\pm$ 0.08 \\
 09-01-2019 & 58492.6229 & 2.16 m/BFOSC & 4000-8700 & -7.2  $\pm$ 0.2 & ...              \\
 27-01-2019 & 58510.5146 & 2.16 m/BFOSC & 5200-8200 & -7.3  $\pm$ 0.1 & ...              \\
 27-01-2019 & 58510.5481 & 2.4 m/YFOSC  & 3500-8750 & -7.0  $\pm$ 0.1 & -0.18 $\pm$ 0.07 \\
 09-02-2019 & 58523.4430 & 2.16 m/BFOSC & 4400-8700 & -6.6  $\pm$ 0.4 & ...              \\
 07-10-2019 & 58763.8718 & 2.16 m/BFOSC & 4000-8700 & -5.5  $\pm$ 0.3 & ...              \\
 03-11-2019 & 58790.7485 & 2.16 m/OMR   & 5500-6900 & -5.2  $\pm$ 0.4 & -0.04 $\pm$ 0.03 \\
 04-11-2019 & 58792.2656 & 2.16 m/OMR   & 5500-6900 & -5.3  $\pm$ 0.3 & 0.01  $\pm$ 0.04 \\
 26-12-2019 & 58843.6260 & 2.4 m/YFOSC  & 4970-9830 & -6.0  $\pm$ 0.2 & 0.06  $\pm$ 0.03 \\
 21-07-2020 & 59052.0035 & SKO/1.3 m    & 5400-7300 & -6.8  $\pm$ 0.1 & -0.01 $\pm$ 0.02 \\
 20-08-2020 & 59082.0787 & SKO/1.3 m    & 5400-7300 & -6.7  $\pm$ 0.0 & -0.01 $\pm$ 0.06 \\
 14-09-2020 & 59107.0080 & SKO/1.3 m    & 5400-7300 & -7.3  $\pm$ 0.1 & ...              \\
 16-09-2020 & 59108.9829 & SKO/1.3 m    & 5400-7300 & -6.7  $\pm$ 0.0 & -0.03 $\pm$ 0.05 \\
 29-09-2020 & 59122.0820 & SKO/1.3 m    & 5400-7300 & -5.7  $\pm$ 0.1 & 0.08  $\pm$ 0.04 \\
 10-10-2020 & 59132.6655 & 2.16 m/BFOSC & 5800-8280 & -6.8  $\pm$ 0.2 & 0.01  $\pm$ 0.05 \\
 11-10-2020 & 59133.8369 & 2.16 m/BFOSC & 5800-8280 & -7.1  $\pm$ 0.1 & 0.07  $\pm$ 0.02 \\
 12-10-2020 & 59134.7450 & 2.16 m/BFOSC & 5800-8280 & -7.0  $\pm$ 0.1 & -0.01 $\pm$ 0.04 \\
 11-08-2021 & 59437.9970 & SKO/1.3 m    & 5400-7300 & -8.0  $\pm$ 0.1 & -0.07 $\pm$ 0.04 \\
 01-09-2021 & 59459.0720 & SKO/1.3 m    & 5400-7300 & -8.3  $\pm$ 0.0 & -0.06 $\pm$ 0.03 \\
 05-09-2021 & 59463.0289 & SKO/1.3 m    & 5400-7300 & -8.2  $\pm$ 0.1 & -0.09 $\pm$ 0.04 \\
 09-10-2021 & 59496.9755 & SKO/1.3 m    & 5400-7300 & -8.2  $\pm$ 0.1 & -0.07 $\pm$ 0.05 \\
 10-10-2021 & 59497.6557 & 2.16 m/OMR   & 6000-6850 & -8.4  $\pm$ 0.1 & -0.11 $\pm$ 0.06 \\
\hline
\end{tabular}
\end{table*}

\section{Table of V\,/\,R ratio and peak separation}

\begin{table*}
\caption{V\,/\,R ratio and peak separation of Swift J0243.6+6124 between 2017 and 2021.}
\label{V_R_table}
 \centering
\begin{tabular}{cclccc}
\hline\hline
Date & MJD & Telescope/ & Wavelength Range & $\log(V\,/\,R)$(H$\alpha$) & $\Delta$V(H$\alpha$)\\
(DD-MM-YYYY) & & Instrument & (\AA) & & (km s$^{-1}$)\\
\hline
 27-10-2017 & 58053.6744 & 2.16 m/OMR   & 5500-6900 &  0.453 & 230\\
 27-10-2017 & 58053.8025 & 2.16 m/OMR   & 5500-6900 &  0.495 & 211\\
 28-10-2017 & 58054.6191 & 2.16 m/OMR   & 5500-6900 &  0.537 & 217\\
 29-10-2017 & 58055.6853 & 2.16 m/OMR   & 5500-6900 &  0.462 & 228\\
 11-11-2017 & 58068.6412 & 2.16 m/OMR   & 5500-6900 &  0.390 & 195\\
 11-11-2017 & 58068.7547 & 2.16 m/OMR   & 5500-6900 &  0.404 & 195\\
 12-11-2017 & 58069.7245 & 2.16 m/OMR   & 5500-6900 &  0.516 & 197\\
 17-09-2018 & 58378.7433 & 2.16 m/OMR   & 5500-6900 &  0.603 & 222\\
 15-11-2018 & 58437.6707 & 2.16 m/BFOSC & 4000-8700 & -0.042 & ...\\
 25-12-2018 & 58477.5579 & 2.4 m/YFOSC  & 4970-9830 & -0.406 & 232\\
 03-11-2019 & 58790.7485 & 2.16 m/OMR   & 5500-6900 & -0.193 & 212\\
 04-11-2019 & 58792.2656 & 2.16 m/OMR   & 5500-6900 & -0.168 & 185\\
 20-08-2020 & 59082.0787 & SKO/1.3 m    & 5400-7300 &  0.022 & 140\\
 09-10-2021 & 59496.9755 & SKO/1.3 m    & 5400-7300 &  0.314 & 218\\
 10-10-2021 & 59497.6557 & 2.16 m/OMR   & 6000-6850 &  0.198 & 207\\
\hline
\end{tabular}
\end{table*}

\section{Table of photometric observations}

\longtab[1]{
\begin{longtable}{cccccc}
\caption{Photometric observations of Swift J0243.6+6124 between 2017 and 2022.}\\
\hline
\hline
 MJD & Telescope & \textit{B} & \textit{V} & \textit{R} & \textit{I} \\
  & & (mag) & (mag) & (mag) & (mag) \\
\hline
\endfirsthead
\caption{continued.} \\
\hline
\hline
 MJD & Telescope & \textit{B} & \textit{V} & \textit{R} & \textit{I} \\
  & & (mag) & (mag) & (mag) & (mag) \\
\hline
\endhead
\hline
\endfoot
\hline
\endlastfoot
58038.62 & 60cm  & 13.725 $\pm$ 0.010 & 12.777 $\pm$ 0.010 & 12.065 $\pm$ 0.010 & 11.241 $\pm$ 0.020 \\
58039.81 & 60cm  & 13.704 $\pm$ 0.010 & 12.729 $\pm$ 0.010 & 12.030 $\pm$ 0.010 & 11.206 $\pm$ 0.020 \\
58041.76 & 60cm  & 13.652 $\pm$ 0.010 & 12.720 $\pm$ 0.010 & 12.014 $\pm$ 0.010 & 11.176 $\pm$ 0.020 \\
58042.72 & 60cm  & 13.700 $\pm$ 0.010 & 12.755 $\pm$ 0.010 & 12.046 $\pm$ 0.010 & 11.234 $\pm$ 0.020 \\
58045.72 & 60cm  & 13.724 $\pm$ 0.010 & 12.768 $\pm$ 0.010 & 12.066 $\pm$ 0.010 & 11.216 $\pm$ 0.020 \\
58046.63 & 60cm  & 13.710 $\pm$ 0.010 & 12.757 $\pm$ 0.010 & 12.036 $\pm$ 0.010 & 11.198 $\pm$ 0.020 \\
58049.86 & 60cm  & 13.681 $\pm$ 0.010 & 12.680 $\pm$ 0.010 & 12.232 $\pm$ 0.010 & 11.152 $\pm$ 0.020 \\
58052.78 & 60cm  & 13.582 $\pm$ 0.010 & 12.632 $\pm$ 0.010 & 11.963 $\pm$ 0.010 & 11.093 $\pm$ 0.020 \\
58053.70 & 80cm  & 13.644 $\pm$ 0.010 & 12.702 $\pm$ 0.010 & 11.974 $\pm$ 0.010 & 11.156 $\pm$ 0.020 \\
58053.77 & 60cm  & 13.636 $\pm$ 0.010 & 12.687 $\pm$ 0.010 & 11.944 $\pm$ 0.010 & 11.131 $\pm$ 0.020 \\
58054.63 & 60cm  & 13.405 $\pm$ 0.010 & 12.476 $\pm$ 0.010 & 11.861 $\pm$ 0.010 & 11.058 $\pm$ 0.020 \\
58054.70 & 80cm  & 13.543 $\pm$ 0.010 & 12.639 $\pm$ 0.011 & 11.942 $\pm$ 0.011 & 11.137 $\pm$ 0.021 \\
58055.64 & 60cm  & 13.617 $\pm$ 0.010 & 12.671 $\pm$ 0.010 & 11.963 $\pm$ 0.010 & 11.155 $\pm$ 0.020 \\
58055.70 & 80cm  & 13.631 $\pm$ 0.010 & 12.683 $\pm$ 0.010 & 11.961 $\pm$ 0.011 & 11.140 $\pm$ 0.020 \\
58056.62 & 60cm  & 13.626 $\pm$ 0.010 & 12.680 $\pm$ 0.010 & 11.977 $\pm$ 0.010 & 11.160 $\pm$ 0.020 \\
58056.74 & 80cm  & 13.629 $\pm$ 0.010 & 12.681 $\pm$ 0.011 & 11.952 $\pm$ 0.011 & 11.126 $\pm$ 0.020 \\
58059.78 & 60cm  & 13.233 $\pm$ 0.010 & 12.361 $\pm$ 0.010 & 11.694 $\pm$ 0.010 & 10.949 $\pm$ 0.020 \\
58061.71 & 80cm  & 13.465 $\pm$ 0.017 & 12.577 $\pm$ 0.015 & 11.874 $\pm$ 0.013 & 11.038 $\pm$ 0.022 \\
58061.80 & 60cm  & 13.443 $\pm$ 0.010 & 12.491 $\pm$ 0.010 & 11.767 $\pm$ 0.010 & 10.956 $\pm$ 0.020 \\
58062.65 & 80cm  & 13.547 $\pm$ 0.014 & 12.600 $\pm$ 0.013 & 11.857 $\pm$ 0.012 & 11.028 $\pm$ 0.021 \\
58063.56 & 80cm  & 13.433 $\pm$ 0.027 & 12.537 $\pm$ 0.023 & 11.801 $\pm$ 0.019 & 10.992 $\pm$ 0.024 \\
58064.65 & 80cm  & 13.483 $\pm$ 0.011 & 12.556 $\pm$ 0.011 & 11.838 $\pm$ 0.011 & 11.007 $\pm$ 0.020 \\
58065.69 & 80cm  & 13.541 $\pm$ 0.012 & 12.574 $\pm$ 0.012 & 11.846 $\pm$ 0.012 & 11.019 $\pm$ 0.021 \\
58067.57 & 60cm  & 13.398 $\pm$ 0.010 & 12.471 $\pm$ 0.010 & 11.804 $\pm$ 0.010 & 10.975 $\pm$ 0.020 \\
58069.57 & 60cm  & 13.504 $\pm$ 0.010 & 12.554 $\pm$ 0.010 & 11.849 $\pm$ 0.010 & 11.008 $\pm$ 0.020 \\
58073.52 & 60cm  & 13.571 $\pm$ 0.010 & 12.613 $\pm$ 0.010 & 11.890 $\pm$ 0.010 & 11.078 $\pm$ 0.020 \\
58074.65 & 80cm  & 13.440 $\pm$ 0.010 & 12.518 $\pm$ 0.011 & 11.814 $\pm$ 0.011 & 10.990 $\pm$ 0.021 \\
58080.61 & 2.4m  & 13.628 $\pm$ 0.010 & 12.663 $\pm$ 0.010 & 11.932 $\pm$ 0.010 & 11.102 $\pm$ 0.020 \\
58081.53 & 2.4m  & 13.638 $\pm$ 0.010 & 12.677 $\pm$ 0.010 & 11.943 $\pm$ 0.010 & 11.103 $\pm$ 0.020 \\
58082.55 & 2.4m  & 13.649 $\pm$ 0.010 & 12.681 $\pm$ 0.010 & 11.942 $\pm$ 0.010 & 11.122 $\pm$ 0.020 \\
58083.60 & 60cm  & 13.540 $\pm$ 0.010 & 12.623 $\pm$ 0.010 & 11.913 $\pm$ 0.010 & 11.096 $\pm$ 0.020 \\
58086.62 & 60cm  & 13.572 $\pm$ 0.010 & 12.630 $\pm$ 0.010 & 11.872 $\pm$ 0.010 & 11.142 $\pm$ 0.020 \\
58088.56 & 60cm  & 13.591 $\pm$ 0.010 & 12.643 $\pm$ 0.010 & 11.931 $\pm$ 0.010 & 11.129 $\pm$ 0.020 \\
58095.62 & 60cm  & 13.661 $\pm$ 0.010 & 12.716 $\pm$ 0.010 & 12.002 $\pm$ 0.010 & 11.187 $\pm$ 0.020 \\
58117.58 & 80cm  & 13.700 $\pm$ 0.011 & 12.762 $\pm$ 0.011 & 12.103 $\pm$ 0.011 & 11.215 $\pm$ 0.020 \\
58118.57 & 80cm  & 13.720 $\pm$ 0.011 & 12.774 $\pm$ 0.011 & 12.040 $\pm$ 0.011 & 11.218 $\pm$ 0.020 \\
58123.64 & 60cm  & 13.733 $\pm$ 0.010 & 12.800 $\pm$ 0.010 & 12.099 $\pm$ 0.010 & 11.296 $\pm$ 0.020 \\
58125.64 & 2.4m  & 13.735 $\pm$ 0.010 & 12.783 $\pm$ 0.010 & 12.047 $\pm$ 0.011 & 11.242 $\pm$ 0.020 \\
58126.56 & 2.4m  & 13.699 $\pm$ 0.010 & 12.772 $\pm$ 0.010 & 12.068 $\pm$ 0.010 & 11.235 $\pm$ 0.020 \\
58135.60 & 60cm  & 13.751 $\pm$ 0.010 & 12.805 $\pm$ 0.010 & 12.097 $\pm$ 0.010 & 11.285 $\pm$ 0.020 \\
58138.57 & 60cm  & 13.488 $\pm$ 0.010 & 12.770 $\pm$ 0.010 & 12.064 $\pm$ 0.010 & 11.119 $\pm$ 0.020 \\
58148.53 & 60cm  & 13.724 $\pm$ 0.010 & 12.794 $\pm$ 0.010 & 12.079 $\pm$ 0.010 & 11.298 $\pm$ 0.020 \\
58150.51 & 60cm  & 13.739 $\pm$ 0.010 & 12.786 $\pm$ 0.010 & 12.088 $\pm$ 0.010 & 11.283 $\pm$ 0.020 \\
58152.52 & 60cm  & 13.625 $\pm$ 0.010 & 12.825 $\pm$ 0.010 & 12.145 $\pm$ 0.010 & ...                \\
58153.52 & 60cm  & 13.732 $\pm$ 0.010 & 12.812 $\pm$ 0.010 & 12.128 $\pm$ 0.010 & 11.320 $\pm$ 0.020 \\
58156.43 & 60cm  & 13.690 $\pm$ 0.010 & 12.803 $\pm$ 0.010 & 12.081 $\pm$ 0.010 & 11.277 $\pm$ 0.020 \\
58158.44 & 60cm  & 13.733 $\pm$ 0.010 & 12.788 $\pm$ 0.010 & 12.085 $\pm$ 0.010 & 11.264 $\pm$ 0.020 \\
58161.47 & 60cm  & 13.751 $\pm$ 0.010 & 12.758 $\pm$ 0.010 & 12.044 $\pm$ 0.010 & 11.219 $\pm$ 0.020 \\
58171.47 & 60cm  & 13.792 $\pm$ 0.010 & 12.845 $\pm$ 0.010 & 12.116 $\pm$ 0.010 & 11.340 $\pm$ 0.020 \\
58173.49 & 60cm  & 13.719 $\pm$ 0.010 & 12.792 $\pm$ 0.010 & 12.103 $\pm$ 0.010 & 11.290 $\pm$ 0.020 \\
58175.47 & 60cm  & 13.756 $\pm$ 0.010 & 12.815 $\pm$ 0.010 & 12.128 $\pm$ 0.010 & 11.325 $\pm$ 0.020 \\
58178.47 & 60cm  & 13.735 $\pm$ 0.010 & 12.847 $\pm$ 0.010 & 12.149 $\pm$ 0.010 & 11.337 $\pm$ 0.020 \\
58179.46 & 60cm  & 13.698 $\pm$ 0.010 & 12.780 $\pm$ 0.010 & 12.111 $\pm$ 0.010 & 11.346 $\pm$ 0.020 \\
58185.46 & 60cm  & 13.779 $\pm$ 0.010 & 12.843 $\pm$ 0.010 & 12.125 $\pm$ 0.010 & 11.323 $\pm$ 0.020 \\
58186.46 & 60cm  & 13.626 $\pm$ 0.010 & 12.763 $\pm$ 0.010 & 12.052 $\pm$ 0.010 & 11.250 $\pm$ 0.020 \\
58187.47 & 60cm  & 13.712 $\pm$ 0.010 & 12.769 $\pm$ 0.010 & 12.114 $\pm$ 0.010 & 11.312 $\pm$ 0.020 \\
58192.46 & 60cm  & 13.727 $\pm$ 0.010 & 12.794 $\pm$ 0.010 & 12.067 $\pm$ 0.010 & 11.296 $\pm$ 0.020 \\
58200.46 & 60cm  & 13.780 $\pm$ 0.010 & 12.844 $\pm$ 0.010 & 12.135 $\pm$ 0.010 & 11.315 $\pm$ 0.020 \\
58201.49 & 60cm  & 13.805 $\pm$ 0.010 & 12.873 $\pm$ 0.010 & 12.184 $\pm$ 0.010 & 11.381 $\pm$ 0.020 \\
58202.48 & 60cm  & 13.771 $\pm$ 0.010 & 12.836 $\pm$ 0.010 & 12.158 $\pm$ 0.010 & 11.372 $\pm$ 0.020 \\
58203.49 & 60cm  & 13.675 $\pm$ 0.012 & 12.757 $\pm$ 0.011 & 12.081 $\pm$ 0.010 & 11.293 $\pm$ 0.020 \\
58216.47 & 60cm  & 13.790 $\pm$ 0.011 & 12.851 $\pm$ 0.010 & 12.136 $\pm$ 0.010 & 11.355 $\pm$ 0.020 \\
58218.48 & 60cm  & 13.738 $\pm$ 0.010 & 12.844 $\pm$ 0.010 & 12.088 $\pm$ 0.010 & 11.365 $\pm$ 0.020 \\
58219.48 & 60cm  & 13.780 $\pm$ 0.010 & 12.860 $\pm$ 0.010 & 12.137 $\pm$ 0.010 & 11.376 $\pm$ 0.020 \\
58368.74 & 60cm  & 13.863 $\pm$ 0.010 & 12.899 $\pm$ 0.010 & 12.275 $\pm$ 0.010 & 11.540 $\pm$ 0.020 \\
58369.76 & 80cm  & 13.827 $\pm$ 0.012 & 12.920 $\pm$ 0.011 & 12.059 $\pm$ 0.010 & 11.504 $\pm$ 0.021 \\
58370.78 & 60cm  & 13.806 $\pm$ 0.010 & 12.905 $\pm$ 0.010 & 12.238 $\pm$ 0.010 & 11.474 $\pm$ 0.020 \\
58377.75 & 80cm  & 13.859 $\pm$ 0.011 & 12.940 $\pm$ 0.010 & 12.259 $\pm$ 0.011 & 11.498 $\pm$ 0.020 \\
58378.79 & 80cm  & 13.854 $\pm$ 0.011 & 12.935 $\pm$ 0.010 & 12.273 $\pm$ 0.010 & 11.481 $\pm$ 0.020 \\
58382.83 & 60cm  & 13.849 $\pm$ 0.010 & 12.926 $\pm$ 0.010 & 12.280 $\pm$ 0.010 & 11.496 $\pm$ 0.020 \\
58383.78 & 60cm  & 13.853 $\pm$ 0.010 & 12.941 $\pm$ 0.010 & 12.295 $\pm$ 0.010 & 11.501 $\pm$ 0.020 \\
58385.75 & 60cm  & 13.803 $\pm$ 0.010 & 12.873 $\pm$ 0.010 & 12.206 $\pm$ 0.010 & 11.510 $\pm$ 0.020 \\
58397.71 & 60cm  & 13.860 $\pm$ 0.010 & 12.960 $\pm$ 0.010 & 12.310 $\pm$ 0.010 & 11.529 $\pm$ 0.020 \\
58398.80 & 60cm  & 13.855 $\pm$ 0.010 & 12.930 $\pm$ 0.010 & 12.332 $\pm$ 0.010 & 11.552 $\pm$ 0.020 \\
58408.80 & 60cm  & 13.887 $\pm$ 0.010 & 12.963 $\pm$ 0.010 & 12.285 $\pm$ 0.010 & 11.519 $\pm$ 0.020 \\
58409.79 & 60cm  & 13.873 $\pm$ 0.010 & 12.945 $\pm$ 0.010 & 12.297 $\pm$ 0.010 & 11.527 $\pm$ 0.020 \\
58410.72 & 60cm  & 13.886 $\pm$ 0.010 & 12.971 $\pm$ 0.010 & 12.282 $\pm$ 0.010 & 11.514 $\pm$ 0.020 \\
58411.67 & 60cm  & 13.812 $\pm$ 0.010 & 12.922 $\pm$ 0.010 & 12.268 $\pm$ 0.010 & 11.494 $\pm$ 0.020 \\
58414.58 & 60cm  & 13.850 $\pm$ 0.010 & 12.951 $\pm$ 0.010 & 12.293 $\pm$ 0.010 & 11.530 $\pm$ 0.020 \\
58421.69 & 60cm  & 13.887 $\pm$ 0.010 & 12.969 $\pm$ 0.010 & 12.294 $\pm$ 0.010 & 11.520 $\pm$ 0.020 \\
58422.67 & 60cm  & 13.919 $\pm$ 0.010 & 12.970 $\pm$ 0.010 & 12.300 $\pm$ 0.010 & 11.525 $\pm$ 0.020 \\
58425.63 & 60cm  & 13.858 $\pm$ 0.010 & 12.977 $\pm$ 0.010 & 12.323 $\pm$ 0.010 & 11.549 $\pm$ 0.020 \\
58428.75 & 80cm  & 13.855 $\pm$ 0.011 & 12.967 $\pm$ 0.010 & 12.268 $\pm$ 0.010 & 11.557 $\pm$ 0.020 \\
58430.68 & 80cm  & 13.883 $\pm$ 0.011 & 12.978 $\pm$ 0.010 & 12.313 $\pm$ 0.010 & 11.564 $\pm$ 0.020 \\
58456.62 & 80cm  & 13.896 $\pm$ 0.011 & 12.993 $\pm$ 0.010 & 12.324 $\pm$ 0.011 & 11.581 $\pm$ 0.021 \\
58468.53 & 60cm  & 13.902 $\pm$ 0.010 & 12.996 $\pm$ 0.010 & 12.334 $\pm$ 0.010 & 11.570 $\pm$ 0.020 \\
58469.59 & 60cm  & 13.896 $\pm$ 0.010 & 12.973 $\pm$ 0.010 & 12.342 $\pm$ 0.010 & 11.571 $\pm$ 0.020 \\
58476.54 & 60cm  & 13.814 $\pm$ 0.010 & 12.921 $\pm$ 0.010 & 12.276 $\pm$ 0.010 & 11.512 $\pm$ 0.020 \\
58477.53 & 60cm  & 13.835 $\pm$ 0.010 & 12.943 $\pm$ 0.010 & 12.285 $\pm$ 0.010 & 11.525 $\pm$ 0.020 \\
58477.55 & 2.4m  & 13.864 $\pm$ 0.010 & 12.963 $\pm$ 0.010 & 12.339 $\pm$ 0.010 & 11.532 $\pm$ 0.020 \\
58478.54 & 60cm  & 13.844 $\pm$ 0.010 & 12.938 $\pm$ 0.010 & 12.275 $\pm$ 0.010 & 11.523 $\pm$ 0.020 \\
58478.55 & 2.4m  & 13.881 $\pm$ 0.010 & 12.971 $\pm$ 0.010 & 12.301 $\pm$ 0.010 & 11.543 $\pm$ 0.020 \\
58479.60 & 2.4m  & 13.884 $\pm$ 0.010 & 12.975 $\pm$ 0.010 & 12.304 $\pm$ 0.010 & 11.548 $\pm$ 0.020 \\
58480.54 & 2.4m  & 13.887 $\pm$ 0.010 & 12.970 $\pm$ 0.010 & 12.306 $\pm$ 0.010 & 11.541 $\pm$ 0.020 \\
58481.55 & 60cm  & 13.806 $\pm$ 0.010 & 12.938 $\pm$ 0.010 & 12.316 $\pm$ 0.010 & 11.516 $\pm$ 0.020 \\
58482.55 & 60cm  & 13.890 $\pm$ 0.010 & 12.982 $\pm$ 0.010 & 12.330 $\pm$ 0.010 & 11.554 $\pm$ 0.020 \\
58484.53 & 60cm  & 13.885 $\pm$ 0.010 & 12.974 $\pm$ 0.010 & 12.320 $\pm$ 0.010 & 11.554 $\pm$ 0.020 \\
58485.54 & 60cm  & 13.876 $\pm$ 0.010 & 12.960 $\pm$ 0.010 & 12.296 $\pm$ 0.010 & 11.538 $\pm$ 0.020 \\
58487.53 & 60cm  & 13.806 $\pm$ 0.010 & 12.894 $\pm$ 0.010 & 12.296 $\pm$ 0.010 & 11.520 $\pm$ 0.020 \\
58489.58 & 60cm  & 13.881 $\pm$ 0.010 & 12.965 $\pm$ 0.010 & 12.303 $\pm$ 0.010 & 11.506 $\pm$ 0.020 \\
58490.63 & 60cm  & 13.797 $\pm$ 0.010 & 12.881 $\pm$ 0.010 & 12.278 $\pm$ 0.010 & 11.498 $\pm$ 0.020 \\
58491.52 & 60cm  & 13.846 $\pm$ 0.010 & 12.965 $\pm$ 0.010 & 12.306 $\pm$ 0.010 & 11.533 $\pm$ 0.020 \\
58499.49 & 60cm  & 13.822 $\pm$ 0.010 & 12.925 $\pm$ 0.010 & 12.287 $\pm$ 0.010 & 11.517 $\pm$ 0.020 \\
58500.46 & 60cm  & 13.847 $\pm$ 0.010 & 12.977 $\pm$ 0.010 & 12.302 $\pm$ 0.010 & 11.557 $\pm$ 0.020 \\
58501.48 & 60cm  & 13.852 $\pm$ 0.010 & 12.951 $\pm$ 0.010 & 12.261 $\pm$ 0.010 & 11.527 $\pm$ 0.020 \\
58504.47 & 60cm  & 13.869 $\pm$ 0.010 & 12.955 $\pm$ 0.010 & 12.313 $\pm$ 0.010 & 11.556 $\pm$ 0.020 \\
58509.44 & 60cm  & 13.899 $\pm$ 0.010 & 12.990 $\pm$ 0.010 & 12.318 $\pm$ 0.010 & 11.567 $\pm$ 0.020 \\
58510.45 & 60cm  & 13.905 $\pm$ 0.010 & 13.006 $\pm$ 0.010 & 12.348 $\pm$ 0.010 & 11.561 $\pm$ 0.020 \\
58510.54 & 2.16m & 13.844 $\pm$ 0.010 & 12.945 $\pm$ 0.010 & 12.301 $\pm$ 0.010 & 11.534 $\pm$ 0.020 \\
58511.46 & 60cm  & 13.904 $\pm$ 0.010 & 12.994 $\pm$ 0.010 & 12.360 $\pm$ 0.010 & 11.584 $\pm$ 0.020 \\
58514.47 & 60cm  & 13.850 $\pm$ 0.010 & 12.957 $\pm$ 0.010 & 12.281 $\pm$ 0.010 & 11.545 $\pm$ 0.020 \\
58515.45 & 60cm  & 13.908 $\pm$ 0.010 & 12.998 $\pm$ 0.010 & 12.370 $\pm$ 0.010 & 11.565 $\pm$ 0.020 \\
58525.45 & 60cm  & 13.826 $\pm$ 0.010 & 12.909 $\pm$ 0.010 & 12.275 $\pm$ 0.010 & 11.538 $\pm$ 0.020 \\
58534.47 & 60cm  & 13.886 $\pm$ 0.010 & 12.997 $\pm$ 0.010 & ...                & 11.597 $\pm$ 0.020 \\
58537.45 & 60cm  & 13.872 $\pm$ 0.010 & 12.963 $\pm$ 0.010 & 12.315 $\pm$ 0.010 & 11.564 $\pm$ 0.020 \\
58538.52 & 60cm  & 13.899 $\pm$ 0.010 & 12.952 $\pm$ 0.010 & 12.318 $\pm$ 0.010 & 11.558 $\pm$ 0.020 \\
58728.83 & 60cm  & 13.835 $\pm$ 0.010 & 12.893 $\pm$ 0.010 & 12.233 $\pm$ 0.010 & ...                \\
58731.83 & 60cm  & 13.846 $\pm$ 0.010 & 12.926 $\pm$ 0.010 & 12.263 $\pm$ 0.010 & 11.491 $\pm$ 0.020 \\
58732.77 & 60cm  & 13.838 $\pm$ 0.010 & 12.911 $\pm$ 0.010 & 12.246 $\pm$ 0.010 & 11.481 $\pm$ 0.020 \\
58733.87 & 60cm  & 13.765 $\pm$ 0.011 & 12.852 $\pm$ 0.010 & 12.253 $\pm$ 0.010 & 11.431 $\pm$ 0.020 \\
58743.73 & 60cm  & 13.815 $\pm$ 0.010 & 12.915 $\pm$ 0.010 & 12.254 $\pm$ 0.010 & 11.471 $\pm$ 0.020 \\
58744.74 & 60cm  & 13.780 $\pm$ 0.010 & 12.873 $\pm$ 0.010 & 12.193 $\pm$ 0.010 & ...                \\
58749.67 & 60cm  & 13.828 $\pm$ 0.010 & 12.911 $\pm$ 0.010 & 12.237 $\pm$ 0.010 & 11.445 $\pm$ 0.020 \\
58750.75 & 80cm  & 13.812 $\pm$ 0.011 & 12.898 $\pm$ 0.010 & 12.312 $\pm$ 0.010 & 11.461 $\pm$ 0.020 \\
58751.76 & 80cm  & 13.862 $\pm$ 0.011 & 12.937 $\pm$ 0.010 & 12.255 $\pm$ 0.010 & 11.495 $\pm$ 0.020 \\
58752.67 & 60cm  & 13.829 $\pm$ 0.010 & 12.899 $\pm$ 0.010 & 12.247 $\pm$ 0.010 & 11.442 $\pm$ 0.020 \\
58757.61 & 60cm  & 13.777 $\pm$ 0.010 & 12.870 $\pm$ 0.010 & 12.209 $\pm$ 0.010 & 11.424 $\pm$ 0.020 \\
58764.59 & 60cm  & 13.776 $\pm$ 0.010 & 12.832 $\pm$ 0.010 & 12.216 $\pm$ 0.010 & 11.379 $\pm$ 0.020 \\
58776.64 & 60cm  & 13.756 $\pm$ 0.010 & 12.824 $\pm$ 0.010 & 12.108 $\pm$ 0.010 & 11.418 $\pm$ 0.020 \\
58777.69 & 60cm  & 13.820 $\pm$ 0.010 & 12.886 $\pm$ 0.010 & 12.229 $\pm$ 0.010 & 11.443 $\pm$ 0.020 \\
58782.70 & 60cm  & 13.820 $\pm$ 0.010 & 12.868 $\pm$ 0.010 & 12.184 $\pm$ 0.010 & 11.421 $\pm$ 0.020 \\
58785.71 & 60cm  & 13.742 $\pm$ 0.010 & 12.846 $\pm$ 0.010 & 12.187 $\pm$ 0.010 & 11.424 $\pm$ 0.020 \\
58786.68 & 60cm  & 13.780 $\pm$ 0.010 & 12.884 $\pm$ 0.010 & 12.201 $\pm$ 0.010 & 11.453 $\pm$ 0.020 \\
58787.68 & 60cm  & 13.835 $\pm$ 0.010 & 12.903 $\pm$ 0.010 & 12.224 $\pm$ 0.010 & 11.455 $\pm$ 0.020 \\
58790.59 & 60cm  & 13.774 $\pm$ 0.010 & 12.858 $\pm$ 0.010 & 12.191 $\pm$ 0.010 & 11.388 $\pm$ 0.020 \\
58790.63 & 80cm  & 13.798 $\pm$ 0.011 & 12.871 $\pm$ 0.011 & 12.122 $\pm$ 0.010 & 11.437 $\pm$ 0.020 \\
58791.66 & 80cm  & 13.798 $\pm$ 0.011 & 12.876 $\pm$ 0.010 & 12.209 $\pm$ 0.011 & 11.441 $\pm$ 0.020 \\
58792.63 & 60cm  & 13.748 $\pm$ 0.010 & 12.827 $\pm$ 0.010 & 12.170 $\pm$ 0.010 & 11.392 $\pm$ 0.020 \\
58793.63 & 60cm  & 13.762 $\pm$ 0.010 & 12.872 $\pm$ 0.010 & 12.200 $\pm$ 0.010 & 11.465 $\pm$ 0.020 \\
58793.83 & 80cm  & 13.731 $\pm$ 0.011 & 12.809 $\pm$ 0.011 & 12.215 $\pm$ 0.010 & 11.436 $\pm$ 0.021 \\
58794.63 & 60cm  & 13.736 $\pm$ 0.010 & 12.842 $\pm$ 0.010 & 12.214 $\pm$ 0.010 & 11.394 $\pm$ 0.020 \\
58805.61 & 60cm  & 13.724 $\pm$ 0.010 & 12.855 $\pm$ 0.010 & 12.171 $\pm$ 0.010 & 11.429 $\pm$ 0.020 \\
58806.61 & 60cm  & 13.821 $\pm$ 0.010 & 12.899 $\pm$ 0.010 & 12.227 $\pm$ 0.010 & 11.446 $\pm$ 0.020 \\
58807.67 & 60cm  & 13.782 $\pm$ 0.011 & 12.825 $\pm$ 0.010 & 12.185 $\pm$ 0.010 & 11.405 $\pm$ 0.020 \\
58843.64 & 2.4m  & 13.800 $\pm$ 0.010 & 12.877 $\pm$ 0.010 & 12.171 $\pm$ 0.011 & 11.397 $\pm$ 0.020 \\
58852.50 & 60cm  & 13.818 $\pm$ 0.010 & 12.902 $\pm$ 0.010 & 12.226 $\pm$ 0.010 & 11.429 $\pm$ 0.020 \\
59098.77 & 80cm  & 13.771 $\pm$ 0.012 & 12.829 $\pm$ 0.011 & 12.190 $\pm$ 0.010 & 11.339 $\pm$ 0.021 \\
59099.79 & 80cm  & 13.766 $\pm$ 0.011 & 12.828 $\pm$ 0.011 & 12.134 $\pm$ 0.011 & 11.339 $\pm$ 0.020 \\
59100.68 & 80cm  & 13.767 $\pm$ 0.011 & 12.828 $\pm$ 0.011 & 12.125 $\pm$ 0.011 & 11.342 $\pm$ 0.020 \\
59101.70 & 80cm  & 13.768 $\pm$ 0.011 & 12.822 $\pm$ 0.011 & 12.140 $\pm$ 0.011 & 11.336 $\pm$ 0.020 \\
59107.75 & 80cm  & 13.735 $\pm$ 0.011 & 12.800 $\pm$ 0.011 & 12.134 $\pm$ 0.011 & 11.331 $\pm$ 0.020 \\
59108.74 & 80cm  & 13.697 $\pm$ 0.011 & 12.772 $\pm$ 0.010 & 12.130 $\pm$ 0.011 & 11.317 $\pm$ 0.020 \\
59109.75 & 80cm  & 13.756 $\pm$ 0.011 & 12.813 $\pm$ 0.010 & 12.125 $\pm$ 0.011 & 11.322 $\pm$ 0.020 \\
59131.71 & 80cm  & 13.762 $\pm$ 0.014 & 12.814 $\pm$ 0.013 & 12.122 $\pm$ 0.012 & 11.325 $\pm$ 0.022 \\
59133.74 & 80cm  & 13.730 $\pm$ 0.011 & 12.797 $\pm$ 0.010 & 12.117 $\pm$ 0.011 & 11.313 $\pm$ 0.020 \\
59134.76 & 80cm  & 13.756 $\pm$ 0.011 & 12.816 $\pm$ 0.011 & 12.127 $\pm$ 0.011 & 11.324 $\pm$ 0.020 \\
59497.78 & 80cm  & 13.743 $\pm$ 0.011 & 12.803 $\pm$ 0.011 & 12.104 $\pm$ 0.011 & 11.308 $\pm$ 0.020 \\
59564.53 & 80cm' & 13.801 $\pm$ 0.010 & 12.812 $\pm$ 0.010 & 12.117 $\pm$ 0.010 & 11.351 $\pm$ 0.020 \\
59565.51 & 80cm' & 13.797 $\pm$ 0.011 & 12.801 $\pm$ 0.010 & 12.114 $\pm$ 0.010 & 11.345 $\pm$ 0.020 \\
59567.51 & 80cm' & 13.800 $\pm$ 0.010 & 12.804 $\pm$ 0.010 & 12.111 $\pm$ 0.010 & 11.346 $\pm$ 0.020 \\
59568.53 & 80cm' & 13.791 $\pm$ 0.010 & 12.793 $\pm$ 0.010 & 12.100 $\pm$ 0.010 & 11.337 $\pm$ 0.020 \\
59569.57 & 80cm' & 13.780 $\pm$ 0.010 & 12.799 $\pm$ 0.010 & 12.103 $\pm$ 0.010 & 11.341 $\pm$ 0.020 \\
59570.54 & 80cm' & 13.797 $\pm$ 0.010 & 12.803 $\pm$ 0.010 & 12.113 $\pm$ 0.010 & 11.350 $\pm$ 0.020 \\
59571.50 & 80cm' & 13.796 $\pm$ 0.010 & 12.809 $\pm$ 0.011 & 12.098 $\pm$ 0.014 & 11.348 $\pm$ 0.020 \\
59572.51 & 80cm' & 13.799 $\pm$ 0.010 & 12.805 $\pm$ 0.010 & 12.107 $\pm$ 0.010 & 11.359 $\pm$ 0.020 \\
59573.71 & 80cm' & 13.771 $\pm$ 0.019 & 12.798 $\pm$ 0.013 & 12.106 $\pm$ 0.013 & ...                \\
59576.64 & 80cm' & 13.803 $\pm$ 0.010 & 12.798 $\pm$ 0.010 & 12.104 $\pm$ 0.010 & 11.333 $\pm$ 0.020 \\
59577.63 & 80cm' & 13.794 $\pm$ 0.010 & 12.800 $\pm$ 0.010 & 12.106 $\pm$ 0.010 & 11.345 $\pm$ 0.020 \\
59578.60 & 80cm' & 13.799 $\pm$ 0.010 & 12.799 $\pm$ 0.010 & 12.103 $\pm$ 0.010 & 11.343 $\pm$ 0.020 \\
59579.63 & 80cm' & 13.797 $\pm$ 0.010 & 12.801 $\pm$ 0.010 & 12.114 $\pm$ 0.010 & 11.342 $\pm$ 0.020 \\
59580.67 & 80cm' & 13.797 $\pm$ 0.010 & 12.799 $\pm$ 0.010 & 12.103 $\pm$ 0.010 & 11.337 $\pm$ 0.020 \\
59581.67 & 80cm' & 13.784 $\pm$ 0.010 & 12.799 $\pm$ 0.010 & 12.106 $\pm$ 0.010 & 11.339 $\pm$ 0.020 \\
59582.62 & 80cm' & 13.797 $\pm$ 0.010 & 12.805 $\pm$ 0.010 & 12.113 $\pm$ 0.010 & 11.345 $\pm$ 0.020 \\
59583.63 & 80cm' & 13.794 $\pm$ 0.010 & 12.795 $\pm$ 0.010 & 12.107 $\pm$ 0.010 & 11.347 $\pm$ 0.020 \\
59584.62 & 80cm' & 13.795 $\pm$ 0.010 & 12.804 $\pm$ 0.010 & 12.110 $\pm$ 0.010 & 11.356 $\pm$ 0.020 \\
59586.61 & 80cm' & 13.799 $\pm$ 0.010 & 12.807 $\pm$ 0.010 & 12.116 $\pm$ 0.010 & 11.357 $\pm$ 0.020 \\
59587.60 & 80cm' & 13.798 $\pm$ 0.010 & 12.805 $\pm$ 0.010 & 12.118 $\pm$ 0.010 & 11.356 $\pm$ 0.020 \\
59588.60 & 80cm' & 13.797 $\pm$ 0.010 & 12.806 $\pm$ 0.010 & 12.113 $\pm$ 0.010 & 11.350 $\pm$ 0.020 \\
59589.61 & 80cm' & 13.793 $\pm$ 0.010 & 12.797 $\pm$ 0.010 & 12.118 $\pm$ 0.010 & 11.352 $\pm$ 0.020 \\
59590.59 & 80cm' & 13.806 $\pm$ 0.010 & 12.809 $\pm$ 0.010 & 12.119 $\pm$ 0.010 & 11.349 $\pm$ 0.020 \\
59591.63 & 80cm' & 13.804 $\pm$ 0.010 & 12.802 $\pm$ 0.010 & 12.109 $\pm$ 0.010 & 11.346 $\pm$ 0.020 \\
59608.67 & 80cm' & 13.773 $\pm$ 0.010 & 12.794 $\pm$ 0.010 & 12.113 $\pm$ 0.010 & 11.347 $\pm$ 0.020 \\
59612.66 & 80cm' & 13.736 $\pm$ 0.011 & 12.776 $\pm$ 0.010 & 12.094 $\pm$ 0.010 & 11.341 $\pm$ 0.020 \\
59617.61 & 80cm' & 13.809 $\pm$ 0.010 & 12.810 $\pm$ 0.010 & 12.121 $\pm$ 0.010 & 11.358 $\pm$ 0.020 \\
59621.64 & 80cm' & 13.788 $\pm$ 0.015 & ...                & 12.089 $\pm$ 0.011 & 11.328 $\pm$ 0.054 \\
59625.63 & 80cm' & 13.787 $\pm$ 0.011 & 12.791 $\pm$ 0.010 & 12.095 $\pm$ 0.010 & 11.333 $\pm$ 0.020 \\
59634.60 & 80cm' & 13.786 $\pm$ 0.011 & 12.792 $\pm$ 0.010 & 12.091 $\pm$ 0.010 & 11.323 $\pm$ 0.020 \\
59638.59 & 80cm' & 13.789 $\pm$ 0.010 & 12.798 $\pm$ 0.010 & 12.116 $\pm$ 0.010 & 11.349 $\pm$ 0.020 \\
59642.57 & 80cm' & 13.793 $\pm$ 0.010 & 12.803 $\pm$ 0.010 & 12.104 $\pm$ 0.010 & 11.343 $\pm$ 0.020 \\
59646.57 & 80cm' & 13.784 $\pm$ 0.011 & 12.781 $\pm$ 0.010 & 12.104 $\pm$ 0.010 & 11.347 $\pm$ 0.020 \\
59650.56 & 80cm' & 13.784 $\pm$ 0.013 & 12.794 $\pm$ 0.011 & 12.109 $\pm$ 0.013 & 11.337 $\pm$ 0.022 \\
59654.55 & 80cm' & 13.778 $\pm$ 0.019 & 12.800 $\pm$ 0.012 & 12.086 $\pm$ 0.012 & 11.337 $\pm$ 0.021 \\
59658.54 & 80cm' & 13.788 $\pm$ 0.013 & 12.806 $\pm$ 0.011 & 12.103 $\pm$ 0.011 & 11.355 $\pm$ 0.020
\label{phot}
\end{longtable}
}

\end{appendix}


\begin{thebibliography}{}

\bibitem[Bikmaev et al.(2017)]{2017ATel10968....1B} Bikmaev, I., Shimansky, V., Irtuganov, E., et al.\ 2017, The Astronomer's Telegram, 10968


\bibitem[Carciofi(2011)]{2011IAUS..272..325C} Carciofi, A.~C.\ 2011, Active OB Stars: Structure, Evolution, Mass Loss, and Critical Limits, 272, 325. doi:10.1017/S1743921311010738

\bibitem[Doroshenko et al.(2018)]{2018A&A...613A..19D} Doroshenko, V., Tsygankov, S., \& Santangelo, A.\ 2018, \aap, 613, A19. doi:10.1051/0004-6361/201732208

\bibitem[Doroshenko et al.(2020)]{2020MNRAS.491.1857D} Doroshenko, V., Zhang, S.~N., Santangelo, A., et al.\ 2020, \mnras, 491, 1857. doi:10.1093/mnras/stz2879

\bibitem[Gaia Collaboration(2022)]{2022yCat.1355....0G} Gaia Collaboration\ 2022, VizieR Online Data Catalog, I/355

\bibitem[Gaia Collaboration et al.(2016)]{2016A&A...595A...1G} Gaia Collaboration, Prusti, T., de Bruijne, J.~H.~J., et al.\ 2016, \aap, 595, A1. doi:10.1051/0004-6361/201629272



\bibitem[Hanuschik et al.(1995)]{1995A&A...300..163H} Hanuschik, R.~W., Hummel, W., Dietle, O., et al.\ 1995, \aap, 300, 163

\bibitem[Harmanec(1983)]{1983HvaOB...7...55H} Harmanec, P.\ 1983, Hvar Observatory Bulletin, 7, 55



\bibitem[Hayasaki \& Okazaki(2004)]{2004astro.ph.12203H} Hayasaki, K. \& Okazaki, A.~T.\ 2004, astro-ph/0412203, Conference Proceeding for the "Interacting Binaries: Accretion, Evolution and Outcome" worshop held in Cefalu' (Italy), July 2004

\bibitem[Huang et al.(2012)]{2012RAA....12.1585H} Huang, F., Li, J.-Z., Wang, X.-F., et al.\ 2012, Research in Astronomy and Astrophysics, 12, 1585. doi:10.1088/1674-4527/12/11/012


\bibitem[Jayasinghe et al.(2019)]{2019MNRAS.486.1907J} Jayasinghe, T., Stanek, K.~Z., Kochanek, C.~S., et al.\ 2019, \mnras, 486, 1907. doi:10.1093/mnras/stz844

\bibitem[Jenke \& Wilson-Hodge(2017)]{2017ATel10812....1J} Jenke, P. \& Wilson-Hodge, C.~A.\ 2017, The Astronomer's Telegram, 10812

\bibitem[Kennea et al.(2017)]{2017ATel10809....1K} Kennea, J.~A., Lien, A.~Y., Krimm, H.~A., et al.\ 2017, The Astronomer's Telegram, 10809

\bibitem[Kong et al.(2022)]{2022arXiv220604283K} Kong, L.-D., Zhang, S., Zhang, S.-N., et al.\ 2022, arXiv:2206.04283

\bibitem[Krimm et al.(2013)]{2013ApJS..209...14K} Krimm, H.~A., Holland, S.~T., Corbet, R.~H.~D., et al.\ 2013, \apjs, 209, 14. doi:10.1088/0067-0049/209/1/14

\bibitem[Liu et al.(2006)]{2006A&A...455.1165L} Liu, Q.~Z., van Paradijs, J., \& van den Heuvel, E.~P.~J.\ 2006, \aap, 455, 1165. doi:10.1051/0004-6361:20064987

\bibitem[Mainzer et al.(2011)]{2011ApJ...731...53M} Mainzer, A., Bauer, J., Grav, T., et al.\ 2011, \apj, 731, 53. doi:10.1088/0004-637X/731/1/53
  

\bibitem[Meegan et al.(2009)]{2009ApJ...702..791M} Meegan, C., Lichti, G., Bhat, P.~N., et al.\ 2009, \apj, 702, 791. doi:10.1088/0004-637X/702/1/791

\bibitem[Mennickent et al.(1997)]{1997A&A...326.1167M} Mennickent, R.~E., Sterken, C., \& Vogt, N.\ 1997, \aap, 326, 1167

\bibitem[Negueruela et al.(2004)]{2004AN....325..749N} Negueruela, I., Steele, I.~A., \& Bernabeu, G.\ 2004, Astronomische Nachrichten, 325, 749. doi:10.1002/asna.200310258

\bibitem[Nesci(2017)]{2017ATel10989....1N} Nesci, R.\ 2017, The Astronomer's Telegram, 10989

\bibitem[Okazaki(1997)]{1997A&A...318..548O} Okazaki, A.~T.\ 1997, \aap, 318, 548




\bibitem[Reig(2011)]{2011Ap&SS.332....1R} Reig, P.\ 2011, \apss, 332, 1. doi:10.1007/s10509-010-0575-8

\bibitem[Reig \& Roche(1999)]{1999MNRAS.306..100R} Reig, P. \& Roche, P.\ 1999, \mnras, 306, 100. doi:10.1046/j.1365-8711.1999.0\\2473.x

\bibitem[Reig et al.(2016)]{2016A&A...590A.122R} Reig, P., Nersesian, A., Zezas, A., et al.\ 2016, \aap, 590, A122. doi:10.1051/0\\004-6361/201628271

\bibitem[Reig et al.(2020)]{2020A&A...640A..35R} Reig, P., Fabregat, J., \& Alfonso-Garz{\'o}n, J.\ 2020, \aap, 640, A35. doi:10.1051/\\0004-6361/202038333

\bibitem[Rivinius et al.(2013)]{2013A&ARv..21...69R} Rivinius, Thomas, Carciofi, Alex C., and Martayan, Christophe, 2013, The Astronomy and Astrophysics Review, 21, 69

\bibitem[Shappee et al.(2014)]{2014ApJ...788...48S} Shappee, B.~J., Prieto, J.~L., Grupe, D., et al.\ 2014, \apj, 788, 48. doi:10.1088/0\\004-637X/788/1/48

\bibitem[Slettebak et al.(1992)]{1992ApJS...81..335S} Slettebak, A., Collins, G.~W., \& Truax, R.\ 1992, \apjs, 81, 335. doi:10.1086/19\\1696

\bibitem[Telting et al.(1994)]{1994A&A...288..558T} Telting, J.~H., Heemskerk, M.~H.~M., Henrichs, H.~F., et al.\ 1994, \aap, 288, 558


\bibitem[Tsygankov et al.(2018)]{2018MNRAS.479L.134T} Tsygankov, S.~S., Doroshenko, V., Mushtukov, A.~A., et al.\ 2018, \mnras, 479, L134. doi:10.1093/mnrasl/sly116


\bibitem[van den Eijnden et al.(2018)]{2018Natur.562..233V} van den Eijnden, J., Degenaar, N., Russell, T.~D., et al.\ 2018, \nat, 562, 233. doi:10.1038/s41586-018-0524-1

\bibitem[Wang et al.(2008)]{2008ApJ...675..626W} Wang, X., Li, W., Filippenko, A.~V., et al.\ 2008, \apj, 675, 626. doi:10.1086/526413



\bibitem[Zhang et al.(2019)]{2019ApJ...879...61Z} Zhang, Y., Ge, M., Song, L., et al.\ 2019, \apj, 879, 61. doi:10.3847/1538-4357/ab22b1

\bibitem[Ziolkowski(2002)]{2002MmSAI..73.1038Z} Ziolkowski, J.\ 2002, \memsai, 73, 1038

\end{thebibliography}
\end{document}